\pdfoutput=1

\documentclass[11pt,twoside,a4paper,cmspaper,final,collab]{cms-tdr}
\def\svnVersion{333b815-D}\def\svnDate{2019/10/08}\def\cmsCernNoTag{CERN-EP-2019-129}\def\cmsCernDate{\today}\def\cmsMessage{"Published in Physical Review D as \href{http://dx.doi.org/10.1103/PhysRevD.100.072001}{\doi{10.1103/PhysRevD.100.072001}.}"}

\begin{document}\cmsNoteHeader{B2G-18-005}

\hyphenation{had-ron-i-za-tion}
\hyphenation{cal-or-i-me-ter}
\hyphenation{de-vices}
\RCS$HeadURL$
\RCS$Id$

\newlength{\cmsFigWidth}

\ifthenelse{\boolean{cms@external}}{\providecommand{\CL}{C.L.\xspace}}{\providecommand{\CL}{CL\xspace}}

\ifthenelse{\boolean{cms@external}}{\setlength\cmsFigWidth{0.49\textwidth}}{\setlength\cmsFigWidth{0.65\textwidth}}

\ifthenelse{\boolean{cms@external}}{\providecommand{\cmsLeft}{upper\xspace}}{\providecommand{\cmsLeft}{left\xspace}}
\ifthenelse{\boolean{cms@external}}{\providecommand{\cmsRight}{lower\xspace}}{\providecommand{\cmsRight}{right\xspace}}

\newcommand{\htakbig}{\ensuremath{\HT^{\mathrm{AK8}}}\xspace}
\newcommand{\htaksml}{\ensuremath{\HT^{\mathrm{AK4}}}\xspace}

\cmsNoteHeader{B2G-18-005}

\title{Search for pair production of vector-like quarks in the fully hadronic final state}

\date{\today}

\abstract{
The results of two searches for pair production of vector-like \PQT or \PQB quarks in fully hadronic final states are presented, using data from the CMS experiment at a center-of-mass energy of 13\TeV.  The data were collected at the LHC during 2016 and correspond to an integrated luminosity of 35.9\fbinv. A cut-based analysis specifically targets the \cPqb{}\PW{} decay mode of the \PQT quark and allows for the reconstruction of the \PQT quark candidates.  In a second analysis, a multiclassification algorithm, the ``boosted event shape tagger,'' is deployed to label candidate jets as originating from top quarks, and \PW{}, \PZ{}, and \PH{}. Candidate events are categorized according to the multiplicities of identified jets, and the scalar sum of all observed jet momenta is used to discriminate signal events from the quantum chromodynamics multijet background. Both analyses probe all possible branching fraction combinations of the \PQT and \PQB quarks and set limits at $95\%{}$ confidence level on their masses, ranging from 740 to 1370\GeV.  These results represent a significant improvement relative to existing searches in the fully hadronic final state.
}

\hypersetup{%
pdfauthor={CMS Collaboration},%
pdftitle={Search for Pair Production of Vector-Like Quarks in the Fully Hadronic Channel},%
pdfsubject={CMS},%
pdfkeywords={CMS, physics, software, computing}}

\maketitle

\section{Introduction}

With the discovery of a light Higgs boson (\PH{}) by the ATLAS and CMS Collaborations in 2012 \cite{atlas_higgs,cms_higgs, cms_higgs_long}, the standard model (SM) is complete as a low-energy effective theory describing all known fundamental particles and their interactions.  However, several questions still remain with the theory, for example, why  the mass of the observed Higgs boson is 125\GeV, whereas quantum loop corrections would be expected to drive the mass up towards the Planck scale.  Many models of new physics beyond the SM predict additional particles that can affect the quantum corrections to the Higgs boson mass and resolve this so-called ``hierarchy problem''.  New states proposed include new particles such as supersymmetric partners of SM particles, or fourth-generation quarks.

Chiral fourth-generation quarks, \PQtpr or \PQbpr, with identical properties to the SM third-generation \cPqt{} and \cPqb{} quarks, but with larger masses, are effectively excluded because of their impact on the Higgs boson production cross section.  However, many models of new physics, such as those predicting a composite Higgs boson \cite{PhysRevD.75.055014,compHiggs,KAPLAN1991259,Dugan:1984hq, BG:2019}, or ``little-Higgs'' models \cite{PhysRevD.69.075002,Matsedonskyi2013}, include fourth-generation particles of a new type, called vector-like quarks (VLQs), labeled \PQT and B, having electric charges of $+2e/3$ and $-1e/3$, respectively.  These VLQs do not obtain their mass via the Higgs boson Yukawa coupling, and will not affect the values of the Higgs boson production cross section or decay width.  Therefore, these are viable search candidates for the LHC experiments, and are predicted to have masses at the \TeVns scale \cite{DeSimone2013}, allowing the hierarchy problem to be resolved. We do not search for the related X and Y particles.

The VLQs are called ``vector-like'' because their left-handed and right-handed chiralities transform under the same $SU(2) \otimes U(1)$ symmetry group of the SM electroweak gauge bosons.  This leads to several decay modes of the VLQs, through charged- and neutral-current interactions.  Although decays to light first- and second-generation quarks are possible, the dominant decay modes of the VLQs are to third-generation SM quarks\cite{PhysRevD.88.094010}.  The possible decay modes of the VLQs to the third-generation quarks are as follows (charge-conjugate modes implied):
\begin{equation}
\begin{aligned}
\PQT &\to \cPqb{}\PW{}, & \PQB &\to \cPqt{}\PW{}, \\
\PQT &\to \cPqt{}\PZ{}, & \PQB &\to \cPqb{}\PZ{}, \\
\PQT &\to \cPqt{}\PH{}, & \PQB &\to \cPqb{}\PH{}.
\label{eq:decays}
\end{aligned}
\end{equation}Specific model assumptions can influence the proportions of these VLQ decay modes.  Both single and pair production of VLQs are possible, with single production dominating at larger VLQ masses ($\approx$2\TeV), while single and pair production rates are comparable for VLQ masses $\approx$1\TeV{}. This analysis considers only the pair production of VLQs.

Both the ATLAS and CMS Collaborations have recently presented searches for pair production of VLQs.  The CMS Collaboration has searched for \PQT and \PQB quarks in the dilepton final state, targeting VLQ decays to \PZ{} bosons \cite{Sirunyan:2018qau}, and excluding \PQT (\PQB) quark masses up to 1280 (1130)\GeV.  An analysis from CMS including single-lepton, dilepton, and multilepton final states \cite{Sirunyan:2018omb} probes all decay modes of the VLQs, and excludes \PQT quark masses in the range 1140--1300\GeV and \PQB quark masses in 910--1240\GeV, depending on the combination of the VLQ branching fractions.  Finally, a CMS result optimized for the \cPqb{}\PW{}\cPqb{}\PW{} channel, using single-lepton final states, excludes \PQT quark masses up to 1295\GeV \cite{Sirunyan:2017pks}.  The ATLAS Collaboration has recently presented a search for VLQ pair production in the fully hadronic channel, with sensitivity to all possible decay modes of the VLQs \cite{Aaboud:2018wxv}.  This analysis most strongly excludes \PQT and \PQB quarks when they decay to Higgs bosons, with mass exclusion limits of 1010\GeV.  The ATLAS Collaboration has also performed a combination of searches utilizing various final states, resulting in mass exclusion limits of up to 1370\GeV \cite{Aaboud:2018pii}.

In this paper, we describe two independent analyses targeting pair production of vector-like
quarks in fully hadronic final states. We first present an analysis that employs a traditional
strategy, utilizing \PW{} boson tagging and \cPqb{} quark tagging algorithms. This analysis specifically targets the \cPqb{}\PW{} decay mode of the \PQT quark, but is used to evaluate sensitivity to all possible decays of the \PQT or \PQB quark, and is referred to as the ``cut-based analysis''. The second analysis uses a novel machine learning
technique to identify and classify different varieties of Lorentz-boosted particles that originate from
VLQ decays. This strategy allows the analysis to target all the decay modes of the \PQT or \PQB quark. We
refer to this analysis as the ``NN (neural network) analysis''.

The cut-based analysis uses dedicated algorithms to identify efficiently jets consistent with \PW{} bosons and the hadronization of \cPqb{} quarks. These algorithms allow the reconstruction of each VLQ \PQT quark present in the event, providing a mechanism to reduce further the contribution of background processes. At least four jets are required to be present, and events are classified according to the number of jets that are identified as being consistent with a \PW{} boson, to obtain signal regions of varying signal purities. The $\HT$ distribution, defined as the scalar sum of jet transverse momenta (\pt{}), is used for signal discrimination in each category.
 The NN analysis uses a neural network algorithm with a multiple-class output to identify jets as consistent with one of six distinct decay topologies from highly boosted particles: top quark, \PW{} boson, \PZ{} boson, Higgs boson, \cPqb{} quark, and light \cPqu{}/\cPqd{}/\cPqs{}/\cPqc{} quark or gluon (denoted ``light jets''). Events with exactly four jets are considered for the analysis, which is the expected final state for fully hadronic decays of VLQ pairs, as seen in Eq.~\ref{eq:decays}. The multiplicities of jets falling into each of the six categories are used to define 126 independent signal regions, in which the value of $\HT$ is used to discriminate signal from the expected background processes.

The main background contribution in these fully hadronic final states comprises multijet events from quantum chromodynamics (QCD) processes. Techniques based on control samples in data are used to predict the expected QCD multijet background yield and $\HT$ shape. In the cut-based analysis, control regions are used to measure QCD multijet background yields and shapes, which are then extrapolated to the signal regions. In the NN analysis, misidentification rates for each of the six categories of jets considered in the multiclassification algorithm are used to predict the level of contribution of multijet events in the signal regions. Each method is validated using samples of observed and simulated events.

The paper has the following structure. Section \ref{sec:detector} provides a description of the CMS detector and trigger system.  The event reconstruction, including jet reconstruction, jet substructure, and the multiclassification algorithm used in the NN analysis, is described in Section \ref{sec:reco}.  The data sets and simulated samples used are presented in Section \ref{sec:samples}.  Information about the definition of the signal and control regions is included in Section \ref{sec:selection}.  The methods employed to predict the QCD multijet background from data for each analysis are explained in Section \ref{sec:background}, and details of the systematic uncertainties affecting the analyses are itemized in Section \ref{sec:syst}. Signal region yields and distributions are given in Section \ref{sec:disc}, and the statistical analysis used to extract the results is described in Section \ref{sec:stats}.  Finally, the results of the two analyses are presented in Section \ref{sec:results}, and a summary is given in the last section.

\section{The CMS detector}
\label{sec:detector}

The central feature of the CMS apparatus is a superconducting solenoid of 6\unit{m} internal diameter, providing a magnetic field of 3.8\unit{T}. Within the solenoid volume are a silicon pixel and strip tracker, a lead tungstate crystal electromagnetic calorimeter (ECAL), and a brass and scintillator hadron calorimeter (HCAL), each composed of a barrel and two endcap sections. Forward calorimeters extend the pseudorapidity ($\eta$) coverage provided by the barrel and endcap detectors. Muons are detected in gas-ionization chambers embedded in the steel flux-return yoke outside the solenoid.

Events of interest are selected using a two-tiered trigger system~\cite{Khachatryan:2016bia}. The first level, composed of custom hardware processors, uses information from the calorimeters and muon detectors to select events at a rate of around 100\unit{kHz} within a time interval of less than 4\mus. The second level, known as the high-level trigger, consists of a farm of processors running a version of the full event reconstruction software optimized for fast processing, and reduces the event rate to around 1\unit{kHz} before data storage.

A more detailed description of the CMS detector, together with a definition of the coordinate system used and the relevant kinematic variables, can be found in Ref.~\cite{Chatrchyan:2008zzk}.

\section{Event reconstruction}
\label{sec:reco}
To reconstruct and identify each individual particle in an event, a ``particle-flow algorithm''~\cite{CMS-PRF-14-001} that uses an optimized combination of information from the various elements of the CMS detector is employed.  The energy of photons is obtained from the ECAL measurement. The energy of electrons is determined from a combination of the electron momentum at the primary interaction vertex as determined by the tracker, the energy of the corresponding ECAL cluster, and the energy sum of all bremsstrahlung photons spatially compatible with originating from the electron track. The energy of muons is obtained from the curvature of the corresponding track. The energy of charged hadrons is determined from a combination of their momentum measured in the tracker and the matching ECAL and HCAL energy deposits, corrected for zero-suppression effects and for the response function of the calorimeters to hadronic showers. Finally, the energy of neutral hadrons is obtained from the corresponding corrected ECAL and HCAL energy.

The reconstructed vertex with the largest value of summed physics-object $\pt^2$ is taken to be the primary proton-proton ($\Pp\Pp$) interaction vertex. Here the physics objects are the jets, clustered using the jet finding algorithm~\cite{Cacciari:2008gp,Cacciari:2011ma} with the tracks assigned to the vertex as inputs, and the associated missing transverse momentum, taken as the negative vector \pt sum of those jets.

The output of the particle-flow algorithm provides a list of particles that are used as inputs to the jet finding algorithm.  Charged hadrons that are not associated with the primary interaction vertex are removed before jet finding to mitigate the effects of additional $\Pp\Pp$ (``pileup'') interactions occurring in the same or neighboring bunch crossings as the interaction of interest.  The anti-\kt clustering algorithm~\cite{Cacciari:2008gp} is used, as implemented in the \textsc{FastJet} software package \cite{Cacciari:2011ma}, to produce two collections of jets, the first obtained with a distance parameter of $R =0.4$ (AK4 jets), and the second obtained with $R = 0.8$ (AK8 jets), where $R$ is the radius of the jet in the $\eta, \phi$ plane (where $\phi{}$ is the azimuthal angle).  The AK8 jets are used to identify the hadronic decays of massive SM particles, including top quarks, and \PW{}, \PZ{}, and Higgs bosons, while the AK4 jets are used to identify other hadronic activity in the event. The cut-based analysis uses both AK4 and AK8 jets, while the NN analysis only uses the AK8 jets for analysis.

The jet momentum is determined as the vectorial sum of all particle momenta in the jet, and is found from simulation to be within 5 to 10\% of the true momentum over the whole \pt spectrum and detector acceptance. Pileup interactions can contribute additional tracks and calorimetric energy deposits, increasing the apparent jet momentum. To mitigate this effect, tracks identified to be originating from pileup vertices are discarded and an offset correction is applied to correct for remaining contributions. Jet energy corrections are derived from simulation studies so that the average measured response of jets becomes identical to that of particle-level jets. In situ measurements of the momentum balance in dijet, photon+jet, \PZ{}+jet, and multijet events are used to determine any residual differences between the observed and simulated jet energy scale, and to derive appropriate corrections~\cite{Khachatryan:2016kdb}. Additional selection criteria are applied to each jet to remove jets potentially dominated by instrumental effects or reconstruction failures.  The jet energy resolution is approximately 15\% at 10\GeV, 8\% at 100\GeV, and 4\% at 1\TeV.

\subsection{Jet substructure}

To identify the hadronic decays of highly Lorentz-boosted objects, including top quarks, and \PW{}, \PZ{}, and \PH{}, jet substructure information provides powerful discrimination from massive jets originating from QCD multijet production.

The mass of the jet itself can discriminate QCD jets from boosted heavy objects.  A grooming algorithm is applied to jet constituents to better estimate the mass of the originating particle of the jet.  In the algorithm used, the constituents of the AK8 jets are reclustered using the Cambridge--Aachen algorithm~\cite{Dokshitzer:1997in,Wobisch:1998wt}. The ``modified mass drop tagger'' algorithm ~\cite{Dasgupta:2013ihk}, also known as the ``soft-drop'' (SD) algorithm, with angular exponent $\beta = 0$, soft cutoff threshold $z_{\mathrm{cut}} < 0.1$, and characteristic radius $R_{0} = 0.8$~\cite{Larkoski:2014wba}, is applied to remove soft, wide-angle radiation from the jet.  The SD mass ($m_{\mathrm{SD}}$) is used to determine the consistency of a jet with a given boosted heavy object.

In addition to the $m_{\mathrm{SD}}$, information about the distribution of particles within the jet can be used for further discrimination.  A quantity called ``$N$-subjettiness'' \cite{Thaler:2010tr, Thaler:2011gf} is used to determine the consistency of a jet with $N$ or fewer subjets.  The $N$-subjettiness values $\tau_N$ are defined as
\begin{equation}
\tau_N = \frac{1}{d_0} \sum_i p_{\mathrm{T},i}\min\left\{\Delta R_{1,i}, \Delta R_{2,i}, \dots, \Delta R_{N,i}\right],
\end{equation}
where the index $i$ refers to each jet constituent, and $\Delta R\equiv \sqrt{\smash[b]{(\Delta\eta)^2 + (\Delta\phi)^2}}$ is the angular distance between a jet constituent and a candidate subjet axis.  The quantity $d_0$ is a normalization constant.  To identify boosted top quarks, the quantity $\tau_{32}\equiv\tau_3/\tau_2$ is used to target the expected three-subjet signature, while for \PW{}, \PZ{}, and Higgs bosons, the quantity $\tau_{21}$ is used because of the expected two-subjet decay topology.

Jets originating from bottom quarks, which hadronize and subsequently decay, are selected with an algorithm to identify and reconstruct displaced vertices, along with their associated tracking information.  Known as the combined secondary vertex algorithm (CSVv2) \cite{BTV-16-002}, it provides several working points of varying efficiencies and misidentification rates.  In the cut-based analysis, the CSVv2 algorithm is applied to AK4 jets using a working point corresponding to a misidentification probability in simulated \ttbar{} events of 0.01 for \cPqu{}/\cPqd{}/\cPqs{}/\cPg{} jets and an efficiency for identifying genuine \cPqb{} jets of approximately 0.63.  In the NN analysis, the CSVv2 algorithm is applied to the subjets of the AK8 jets to increase the categorization efficiency for decays of top quarks, \PZ and Higgs bosons, which can have one or more displaced vertices within the jet. A CSVv2 working point is not explicitly used in the NN analysis, however the output value of the CSVv2 discriminator for each subjet is used as an input to the multiclassification algorithm to categorize jets.

In the cut-based analysis, a working point for identifying merged decay products of a highly-boosted \PW{} boson in a single jet (\PW{} tagging) is chosen.  To be considered for \PW{} tagging, an AK8 jet must have $\pt > 200$\GeV.  The jet must satisfy $65 < m_{\mathrm{SD}} < 105$\GeV and $\tau_{21} < 0.55$ to be \PW{} tagged.  This working point corresponds to an efficiency of about 0.50 to identify genuine \PW{} jets and a misidentification probability of about 0.03~\cite{CMS:2017wyc}. Because of an observed dependence of $m_{\mathrm{SD}}$ on the \PW{} jet momentum, an additional correction is applied to ensure the \PW{} tagged jet $m_{\mathrm{SD}}$ peak is stable and the \PW{} tagging efficiency remains roughly constant as a function of jet momentum.

\subsection{Boosted event shape tagger (BEST) algorithm}

The NN analysis does not focus on a single VLQ decay mode and thus the expected signatures can contain various combinations of top and bottom quarks along with \PW{}, \PZ{}, and \PH{}.  Using standard cut-based working points for each type of particle leads to complications with overlaps in selection criteria when considering many different final states simultaneously.  For this reason, a new algorithm is used that simultaneously attempts to identify six categories of jets: \cPqt{}, \PW{}, \PZ{}, \PH{}, \cPqb{}, and light jets.  The algorithm is called the boosted event shape tagger (BEST) algorithm, as first detailed in Ref.~\cite{BEST_PRD}, and uses hypothesized reference frames to determine the consistency of a jet with the expected topology from top quark, \PW{}, \PZ{}, \PH{} decays, \cPqb{} quark and light jets.  The algorithm uses a neural network to classify jets according to one of those six possibilities.  The NN analysis presented here is the first CMS result to use the BEST algorithm.

The BEST algorithm relies on the fact that jets from very high energy (``highly boosted'') heavy-particle decays will have a distinct topology in the rest frame of the decaying object.  For example, the decay of a highly boosted \cPqt{} quark produces three collimated particles in the laboratory frame, but in the rest frame of the \cPqt{} quark, the three distinct jet directions lie in a plane.  By Lorentz-boosting the particles or constituents in a jet back to the rest frame, it can be seen whether the distribution of particles is consistent with that expected from a top quark decay.  This boost transformation is applied four different times to obtain four sets of jet constituents. The boost transformation is performed assuming the jet originates from a top quark, \PW{}, \PZ{}, or \PH{}, after forming the boost vector by using the jet four-vector with the mass altered to be that of the particle under consideration, while keeping the jet momentum constant.

The sets of jet constituents resulting from each boost transformation are used to compute kinematic quantities, including Fox--Wolfram moments \cite{PhysRevLett.41.1581}, aplanarity, sphericity, and isotropy, based on the eigenvalues of the sphericity tensor \cite{sphericity}, and the jet thrust \cite{thrust}.  In each boosted reference frame, jet constituents are reclustered to obtain a set of objects relative to the transformed jet axis.  These objects are used to compute the longitudinal asymmetry, defined as the ratio of the longitudinal-component sum of the momenta to the \pt sum of this set of objects.  This ratio gives another way to compute the isotropy of constituents that is expected for a jet consistent with one of the hypothesized particles.   Additionally, the jet $m_{\mathrm{SD}}$, jet $\eta$, charge, $\tau_{32}$, $\tau_{21}$, and subjet CSVv2 scores from the original jet reference frame are used.  In total, 59 kinematic quantities from the original and transformed sets of constituents are used as inputs to a deep neural network to discriminate between the different jet species.  These kinematic quantities are validated by examining distributions in data and simulated events, where good agreement in shape is observed.

The BEST neural network is trained using samples of simulated AK8 jets that originate from the decay of heavy resonances and that correspond to the final state objects (\cPqt{}, \PW{}, \PZ{}, \PH{}, \cPqb{}, or light jets).  The jets in the training sample are matched to the object of interest using the generator-level information.  Samples with heavy resonance masses from 1 to 4\TeV are used to populate the jet \pt range from 0.4 to $\approx$2\TeV.  The neural network is trained using the Python-based \textsc {scikit-learn} package, using the \textsc {MLPClassifier} module \cite{scikitlearn}.  The network architecture consists of 3 hidden layers with 40 nodes in each layer using a rectified-linear activation function.  There are six output nodes, corresponding to the six particle species of interest.  A sample of 500\,000 jets is used to train the network, split evenly between the six training samples. The six outputs from the network represent probabilities for the jet to originate from the corresponding particle.  The classification of an AK8 jet is chosen according to the output node with the highest probability.  Several validation studies have been performed in different samples of data events enriched in different types of processes: a muon+jets sample containing boosted top quarks and boosted \PW{} bosons, a sample containing events from QCD processes enriched in gluon-initiated jets, and a sample of photon+jets events enriched in quark-initiated jets.  In each of these samples, we find good agreement in the shape and rate of the BEST neural network inputs, as well as the output probabilities~\cite{CMS-PAS-JME-18-002}.

\section{Data set and simulated samples}
\label{sec:samples}

Both the cut-based and NN analyses use the data set collected by the CMS experiment at the CERN LHC in 2016, corresponding to an integrated luminosity of $\Pp\Pp$ collisions of 35.9\fbinv.  Events in the cut-based analysis are selected online using a trigger algorithm requiring an $\HT$ value of at least 800\GeV, or 700\GeV if a jet with mass above 50\GeV is present.  Events are also selected by another two triggers, which require a single jet with either $\pt \geq 450$  or 360\GeV with a mass above 30\GeV.  The above trigger selection is measured to be fully efficient for the signal regions, with corrections applied for percent-level inefficiencies in control regions.  Events in the NN analysis are selected online using the above trigger algorithms in combination with all other algorithms requiring multijet topologies.  The trigger requirements for the NN analysis are fully efficient in the signal and control regions, because of the higher jet momenta considered.

Methods utilizing data are employed to estimate the dominant background from QCD multijet production, however, samples of simulated events are used to validate the background estimation techniques described in Section \ref{sec:background}. These samples of QCD multijet events are generated at leading order with \PYTHIA{}~\cite{Sjostrand:2007gs,Sjostrand:2014zea}.

Simulated events are used to model the subdominant background contributions.  The largest of these in both analyses comes from the SM pair production of top quarks, generated at next-to-leading order (NLO) with \POWHEG v2 \cite{Nason:2004rx,Frixione:2007vw} and showered with \PYTHIA 8.212, using the event tune CUETP8M2T4 \cite{pythia_tune}.  The production of a \PW{} or \PZ{} boson in association with additional jets, where the \PW{}/\PZ{} boson decays to quarks, is generated at leading-order (LO) with \MGvATNLO 2.2.2 \cite{Alwall:2014hca,Frederix:2012ps} and also showered with \PYTHIA 8.212.  Diboson events (\PW{}\PW{}, \PW{}\PZ{}, \PZ{}\PZ{}) are generated at LO with \PYTHIA, and rare top quark production processes (\ttbar{}\PW, \ttbar{}\PZ, \ttbar{}\ttbar) are generated at NLO with \MGvATNLO and showered with \PYTHIA.  Background contributions from Higgs boson production in the dominant gluon fusion mode with decays to \bbbar and \PWp\PWm are included via events generated with \MGvATNLO plus \PYTHIA and \POWHEG v2 + \PYTHIA, respectively.  Backgrounds other than \ttbar using \PYTHIA use the CUETP8M1 event tune \cite{pythia_tune2}.  The cut-based analysis considers only the \ttbar and \PW{}+jets background contributions.  Other processes such as \PZ{}+jets were measured to contribute at only the $1\%$ level to the total background expectation, and therefore were not further investigated.

Event samples of pair-produced vector-like \PQT and \PQB quarks, with masses ranging from 0.7 to 1.8\TeV in increments of 100\GeV, are generated at LO using \MGvATNLO \cite{Alwall2011} + \PYTHIA. They are inclusive with respect to the VLQ decay mode, and are generated with equal branching fractions for \PQT/\PQB quark decays to each of the three modes (\cPqt{}\PH{}/\cPqb{}\PH{}, \cPqt{}\PZ{}/\cPqb{}\PZ{}, \cPqb{}\PW{}/\cPqt{}\PW{}).  Events are weighted to produce results for different combinations of branching fractions, and are normalized to theoretical cross section expectations calculated at the next-to-next-to-leading order (NNLO), including next-to-leading-logarithmic order soft-gluon resummation, with \textsc {Top++2.0} \cite{CZAKON20142930}, as listed in Table~\ref{signalXS}.

\begin{table}
\topcaption{Theoretical cross sections for \PQT{}\PQT and \PQB{}\PQB production, calculated at NNLO with \textsc {Top++2.0}.}
\begin{center}
\begin{scotch}{cc}
\PQT/\PQB mass [\GeVns{}] & Cross section [fb] \\
\hline
700 & 455 \\
800 & 196 \\
900 & 90.4 \\
1000 & 44.0 \\
1100 & 22.0 \\
1200 & 11.8 \\
1300 & 6.4 \\
1400 & 3.5 \\
1500 & 2.0 \\
1600 & 1.15 \\
1700 & 0.67 \\
1800 & 0.39 \\
\end{scotch}
\label{signalXS}
\end{center}
\end{table}

\section{Event selection}
\label{sec:selection}

In this section, the event selection and reconstruction techniques applied to the two analyses are described.

\subsection{Cut-based analysis}

The cut-based analysis, optimized for both \PQT quarks decaying to a \cPqb{} quark and \PW{} boson, requires at least two AK8 jets with $\pt > 200$\GeV and $\abs{\eta} < 2.4$.  The AK8 jets serve as boosted \PW{} boson candidates, and are evaluated with the \PW{} boson tagging algorithm described above.  In addition, the analysis requires at least two AK4 jets with $\pt > 30$\GeV and $\abs{\eta} < 2.4$, serving as \cPqb{} jet candidates.  At least two of the selected AK4 jets must be distinct from the AK8 jets, requiring an angular separation of $\Delta R > 0.8$.  If there are more than two AK8 or AK4 jets, the two with the highest \pt are used.  The analysis requires the scalar sum of AK4 jet energies, \htaksml, to be larger than 1200\GeV.  For a signal mass of 1200\GeV, this selection is $95\%$ efficient. With the two AK4 and two AK8 jets, there are two possible combinations of a \PW{} jet and \cPqb{} jet candidate that can be formed.  As signal events are expected to produce two particles with equal mass, we can form the variable
\begin{equation}
\Delta m = 2 \frac{m_{\PQT1} - m_{\PQT2}}{m_{\PQT1} + m_{\PQT2}},
\end{equation}
where $m_{\PQT1}$ is the mass of the higher-\pt \PQT quark candidate and $m_{\PQT2}$ the mass of the lower-\pt \PQT quark candidate, of two \PQT candidates each formed from one AK8 jet and one AK4 jet.  The assignment of AK4 and AK8 jets to \PQT quark candidates is chosen to minimize the value of $\Delta m$.  Events are required to have $\Delta m < 0.1$.

Events passing the \htaksml and $\Delta m$ requirements are further divided into categories.  Applying the \PW{} boson tagging and \cPqb{} quark tagging working points described above, events are divided into categories based on the multiplicity of \PW{} and \cPqb{} tags in the event.  There are nine tagging combinations, with possibilities of 0, 1, or $\geq$2 \PW{} tags in combination with 0, 1, or $\geq$2 \cPqb{} tags.

\subsection{Neural network analysis}

In the NN analysis, each jet in events with exactly four jets is classified according to one of the six categorizations from the BEST
algorithm: \cPqt{}, \cPqb{}, \PW{}, \PZ{}, \PH{}, or light. The number of jets with each BEST classification label is
used to divide events into exclusive categories of varying signal and background contributions,
with categories containing larger numbers of \cPqt{}, \cPqb{}, \PW{}, \PZ{}, or \PH{} candidates being enriched in the
VLQ signal, as it is expected to decay to multiple highly boosted massive objects. In each category, the distribution of \htakbig, which is the scalar sum of the four selected AK8 jet energies, is used to discriminate signal from the background processes.

The signal regions are defined as follows:
\begin{itemize}
\item exactly 4 AK8 jets, each with $\pt{} > 400$\GeV and $\abs{\eta{}} < 2.4$;
\item a unique set of ($N_{\cPqt{}}$, $N_{\PH{}}$, $N_{\PW{}}$, $N_{\PZ{}}$, $N_{\cPqb{}}$, $N_{\mathrm{j}}$), labeling each event by the combination of jet tags.
\end{itemize}

The possible combinations of $N_{\mathrm{i}}$ satisfying the above conditions give 126 independent signal region categories. In some categories, where there is a lack of simulated events to model the subdominant background processes, a single bin is used as a counting experiment instead of the full \htakbig shape information. This occurs in 14 of the 126 total categories, which are:
\parbox{\linewidth}{0\cPqt{}0\PW{}0\PZ{}1\PH{}3\cPqb{}, 0\cPqt{}0\PW{}0\PZ{}2\PH{}2\cPqb{}, 0\cPqt{}0\PW{}2\PZ{}1\PH{}0\cPqb{}, 0\cPqt{}0\PW{}3\PZ{}0\PH{}0\cPqb{}, 0\cPqt{}1\PW{}0\PZ{}0\PH{}3\cPqb{}, 0\cPqt{}1\PW{}0\PZ{}2\PH{}1\cPqb{}, 0\cPqt{}2\PW{}1\PZ{}0\PH{}1\cPqb{}, 1\cPqt{}0\PW{}1\PZ{}2\PH{}0\cPqb{}, 1\cPqt{}0\PW{}2\PZ{}1\PH{}0\cPqb{}, 1\cPqt{}0\PW{}3\PZ{}0\PH{}0\cPqb{}, 1\cPqt{}1\PW{}0\PZ{}2\PH{}0\cPqb{}, 1\cPqt{}3\PW{}0\PZ{}0\PH{}0\cPqb{}, 2\cPqt{}0\PW{}0\PZ{}2\PH{}0\cPqb{}, 2\cPqt{}0\PW{}2\PZ{}0\PH{}0\cPqb{}. No further selections are applied on the jet kinematic variables or the BEST algorithm output probabilities.}

\section{Background estimation methodology}
\label{sec:background}

After the requirements described above have been applied to select the expected signal events, both the cut-based and NN analyses remain dominated by background events from QCD multijet production processes.  Since simulated QCD multijet events do not reliably model the observed data, because of missing higher-order contributions during event generation, both analyses incorporate a method to estimate the background contribution from QCD multijet production directly from observed data events.  This section describes the methodology employed by each analysis. The non-QCD background contributions are taken from simulation.

\subsection{Cut-based analysis}

The cut-based analysis uses an ``ABCD'' matrix method based on observed distributions of two uncorrelated event quantities to predict the shape and rate for the expected QCD multijet background in the signal region.  The two quantities used to define the control regions are \htaksml and $\Delta m$.  The shape of the expected QCD multijet background is obtained by selecting data events passing the $\htaksml > 1200$\GeV requirement, but failing the $\Delta m < 0.1$ selection.  This control region is labeled region B. The expected backgrounds from \ttbar and \PW{}+jets events, as estimated from simulation, are subtracted from the observed distribution to obtain the expected contribution solely from QCD multijet events.  After obtaining the shape, the rate can be estimated by defining another set of control regions, namely with $\htaksml < 1200$\GeV.   This sideband, with $\Delta m < 0.1$, is labeled region A. The ratio of the number of events in A to \PQB (again after subtracting the \ttbar and \PW{}+jets component) results in an extrapolation scale factor.  The control region with $\htaksml{} > 1200$\GeV, $\Delta m > 0.1$ is labeled D. The scale factor is then applied to the shape obtained from region D to describe the expected \htaksml distribution of QCD multijet events in the signal region, labeled region C in this description.

The above procedure is only valid if the quantities \htaksml and $\Delta m$ are uncorrelated.  In simulated QCD multijet events, a small correlation ($<$5\%) is observed, therefore a residual correction is derived from these events.  Specifically, the ABCD procedure is performed in the simulated sample, and the resulting prediction is compared with the observed yield of simulated events in the signal region. A trend in the ratio of these two \htaksml distributions is observed and fit using a linear function.  This function is used to scale the resulting \htaksml distribution in data.  Three functions are derived, for simulated events with 0, 1, or 2 \PW{}-tagged jets. The procedure is validated by applying it in observed events with exactly 0 \PW{}-tagged jets, and agreement is found within 2.5\%.

\subsection{The NN analysis}
\label{sec:qcd}
\begin{figure*}
\begin{center}
\includegraphics[width=0.75\textwidth]{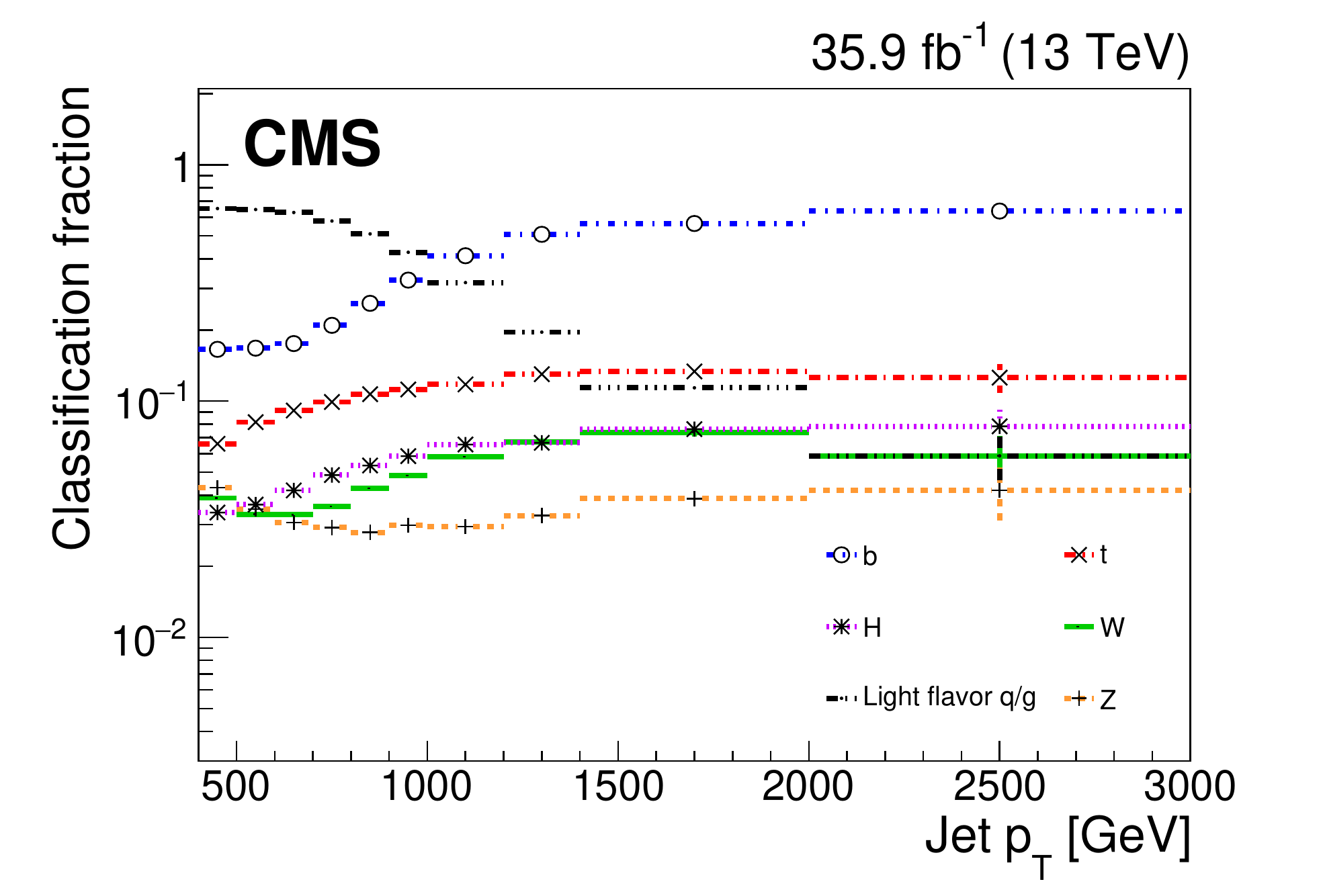}
\caption{Classification fractions for the six categories of the BEST algorithm, measured in data events with exactly three AK8 jets, as a function of jet \pt.  Error bars shown indicate statistical uncertainties in the fractions to be propagated to the estimate of the QCD multijet background contribution.  The rightmost bin includes jets with \pt{} values above 3\TeV.}
\label{fig:BESTrates}
\end{center}
\end{figure*}

The NN analysis uses a method based on the classification fractions of the BEST algorithm to estimate the shape and rate of the QCD multijet background using data.  In the inclusive sample of observed events with exactly three AK8 jets, independent from the four jet sample in which signal is extracted, the classification fraction $\epsilon_X$ for a given jet category $X$ of the BEST algorithm is computed according to the following definition:
\begin{equation}
\epsilon_X ( \pt ) = \frac{N_X}{N},
\end{equation}
where $N_X$ represents the number of jets in BEST category $X$, and $N$ represents the total number of jets.  The classification fraction is measured as a function of jet \pt using data events.  There is negligible signal contamination in this region, which is dominated by QCD multijet events.  The fractions for each BEST category are shown in Fig. \ref{fig:BESTrates}.  These fractions are used to estimate the yield of events having any arbitrary combination of BEST labels.

To obtain the QCD multijet yield as well as the \htakbig shape, the inclusive sample of events with exactly four AK8 jets is used in data, however, the BEST labels are not utilized.  For each of the 126 signal regions in the NN analysis, every event is evaluated as a candidate for the given signal region. Each possible way of assigning jet labels to get the event category is considered, with each assignment resulting in a jet weight according to the classification fraction measured above, as a function of each jet's \pt.  The four jet weights are then multiplied to obtain the final event weight.  The \htakbig value of an event enters the binned \htakbig distribution with the event-level weight applied. After repeating this process for all possible assignments of the BEST labels, and iterating over all events, the final \htakbig distribution for the expected QCD multijet contribution is obtained:
\begin{equation}
r = \sum_{\text{events}}\left\{\sum_{\text{perms}}\left[\prod_{i=1}^4 \epsilon_{X_i} (\pt{}(i))\right]\right\},
\end{equation}
where $r$ represents the expected QCD multijet shape distribution and yield, and the index $i$ corresponds to one of the four jets in the event.

The procedure above is validated in a sample of simulated QCD multijet events, and agreement is obtained between predicted and observed events in both the yield and shape of the \htakbig distribution, within the uncertainties propagated from the measurement of the classification fractions, for all of the 126 signal regions considered.

\section{Systematic uncertainties}
\label{sec:syst}

Several sources of systematic uncertainty are evaluated and included in the final analysis results.  Table \ref{tab:systematics} summarizes the different contributions, and the analysis to which they contribute.  They are described in detail below.

\begin{table*}
\topcaption{Sources of systematic uncertainties that affect the \htaksml or \htakbig distribution in each analysis. Systematic sources with an uncertainty of ``$\pm 1\sigma$'' affect the shape and rate, all others affect the rate only. Sources of systematic error that affect ``all simulation'' impact both the signal simulation and simulated backgrounds.}
\begin{center}
\begin{footnotesize}
\begin{scotch}{lc cc c}
\multicolumn{2}{c}{Uncertainty} & \multicolumn{2}{c}{Contribution to:} &  \\
Source & Uncertainty & Cut-based  & NN &  Applies to samples: \\
\hline
Diboson cross section                & 50\%                   & &\checkmark &  VV only  \\
Rare top quark process cross sections      & 50\%                   & &\checkmark & \ttbar{}V , \ttbar{}\ttbar \\
Higgs boson cross section                  & 50\%                   & &\checkmark & H only \\
\PW{}+jets cross section                 & 15\%                   &\checkmark &\checkmark & \PW{}+jets only \\
\PZ{}+jets cross section                 & 15\%                   & &\checkmark &\PZ{}+jets only \\
Integrated luminosity measurement               & 2.5\%                  &\checkmark &\checkmark & All simulation\\
Pileup reweighting                   & $\pm1\sigma$           &\checkmark &\checkmark & All simulation \\
Jet energy scale                     & $\pm1\sigma(\pt,\eta)$ &\checkmark &\checkmark & All simulation \\
Jet energy resolution                & $\pm1\sigma(\eta)$     &\checkmark &\checkmark & All simulation \\
Parton distribution functions        & $\pm1\sigma$           & \checkmark & \checkmark  & \ttbar, VLQ \\
Renormalization and factorization scales                          & $\pm1\sigma$           &\checkmark &\checkmark & \ttbar, VLQ \\
CSVv2 discriminant reshaping         & $\delta$(wgt., unwgt.) & & \checkmark & All simulation \\
BEST classification fractions                   & $\pm1\sigma(\pt)$           & &\checkmark & QCD multijet \\
BEST classification scale factor                          & 5\%                    & &\checkmark & All simulation \\
BEST misclassification scale factor                       & 5\%                    & &\checkmark & All simulation \\
Trigger                              & 2\%                    & \checkmark & & All simulation \\
\PW{} tag scale factor                   & $\pm1\sigma$           & \checkmark & & All simulation \\
Soft drop jet mass scale              & $\pm1\sigma$           & \checkmark & & All simulation \\
Soft drop jet mass resolution         & $\pm1\sigma$           & \checkmark & & All simulation \\
\cPqb{} tag scale factor                   & $\pm1\sigma$           & \checkmark & & All simulation \\
Extrapolation fit                      & $\pm1\sigma$           & \checkmark & & Background from data \\
Normalization of $1$\PW background prediction     & 1.9\%{}           & \checkmark & & Background from data \\
Normalization of $2$\PW background prediction     & 1.1\%{}           & \checkmark & & Background from data \\

\end{scotch}
\label{tab:systematics}
\end{footnotesize}
\end{center}
\end{table*}

\begin{itemize}

\item {Process cross sections}:  Uncertainties in the cross sections used to normalize simulated background processes are included.  For the \PW{}+jets and \PZ{}+jets backgrounds, uncertainties of 15\% are applied \cite{PhysRevD.96.072005, Sirunyan2018-zxs}.  For the subdominant diboson, rare top quark process (\ttbar{}V, \ttbar{}\ttbar), and Higgs boson contributions, a conservative uncertainty in the cross section value of 50\% is applied.  For \ttbar backgrounds, the uncertainty in the cross section value is included through the scale uncertainties described below, which cover both shape and normalization effects.

\item {Integrated luminosity}:  The uncertainty in the measurement of the integrated luminosity recorded during the 2016 data-taking period by CMS is 2.5\%, and is applied to all simulated signal and background samples \cite{CMS:2017sdi}.

\item {Pileup reweighting}:  All simulated samples used in the analysis are reweighted to ensure the distribution of the number of pileup interactions per event matches the corresponding observed distribution for the 2016 run.  This pileup distribution is obtained using a proton-proton inelastic cross section value of 69.2\unit{mb} \cite{Sirunyan:2018nqx,Aaboud:2016mmw}.  A systematic uncertainty in the distribution is obtained by varying the value by $\pm 4.6$\%, resulting in an uncertainty with both a normalization and shape component.

\item {Jet energy scale and resolution}:  Uncertainties in the corrections applied to jets are propagated to the final discriminating distributions by reconstructing events with the jet-level corrections shifted within their corresponding uncertainties, which depend on the jet \pt and $\eta$ \cite{Khachatryan:2016kdb}.

\item {Parton distribution functions}:  For the \ttbar and VLQ signal simulated samples, we use PDFs from the NNPDF3.0 set \cite{Ball:2014uwa}, and evaluate the systematic uncertainty due to the choice of PDF according to the process described in Ref.~\cite{Butterworth:2015oua}.   For the signal samples, changes in the shape and normalization are considered in the NN-based analysis.  In the cut-based analysis, we find the shape component to be negligible, and consider only a normalization uncertainty.

\item {Scale uncertainties}:  For the \ttbar and VLQ signal simulated samples, we vary the renormalization and factorization scales up and down independently by factors of 2 to account for uncertainties in the choice of scales used to generate the simulated sample.  For the \ttbar samples, the effect associated with this scale variation is sufficiently large to cover the uncertainty in the cross section as well. For the signal samples, changes in the shape and normalization are considered in the NN-based analysis. In the cut-based analysis, we find the shape component to be negligible, and consider only a normalization uncertainty.

\item {The CSVv2 discriminant reshaping (NN-based)}: When using the shape of the CSV discriminant, as we do for inputs to the BEST algorithm, a reshaping event weight is applied based on the CSVv2 scores of the AK8 jets \cite{BTV-16-002}.  We keep the nominal analysis result without the addition of these CSVv2 reshaping weights, but add an additional systematic uncertainty where the standard deviation (s.d.) value is the difference between applying the weights and not applying the weights.

\item {The BEST classification scale factors (NN-based)}:  Uncertainties in the classification and misclassification scale factors are included through 11 independent nuisance parameters, one each for the classification and misclassification efficiencies for the 5 heavy objects (\cPqt{}, \PW{}, \PZ{}, \PH{}, \cPqb{}), and a final nuisance for the QCD categorization efficiency .  Weights are applied on a jet-by-jet basis in each event to produce shape variations in each of the signal regions.  An uncertainty of 5\% per BEST classification is used to compute event weights, and shape templates are formed for each category of the BEST algorithm, separately for correctly and incorrectly classified jets.  This uncertainty is allowed to float during the signal extraction, to measure a value for each scale factor.

\item {The BEST classification fractions for data-driven method (NN-based)}:  We propagate the uncertainty in the measurement of the classification fractions, due to limited event counts in data control regions, to the background estimate.  The uncertainties from the six classification fractions $\epsilon_X$ used are added in quadrature to obtain the total uncertainty for a given event of the expected QCD multijet background distribution $r$, as described in Section \ref{sec:qcd}.

\item{Trigger uncertainty}: We measure the trigger efficiency to be \textgreater{} $99\%$ in our signal region.  A 2$\%$ uncertainty is applied to cover small observed trigger inefficiencies for events with low \htaksml values.  The impact of the trigger inefficiency has been measured to be negligible for the NN analysis because the signal regions are higher in jet momenta and therefore further away from the trigger turn-on region. No additional systematic uncertainty is applied to simulated events in the NN analysis.

\item {\PW{} tagging scale factor uncertainty (cut-based)}: We apply scale factors to account for the difference in \PW{} jet tagging efficiency between simulation and data.  The factor is applied as a weight to simulated events based on the number of \PW{} tags.  The uncertainty in this factor is $14\%$, plus a small factor due to extrapolating the tagging efficiency to higher \pt. The uncertainty for each tag is increased by $4.1\% \log {\pt}_{\PW{}}/200$, where ${\pt}_{\PW{}}$ is the transverse momentum of the tagged \PW{} jet.

\item {Soft drop jet mass scale and resolution (cut-based)}: To account for the uncertainty in the soft drop selection used in \PW tagging, the jet $m_{\mathrm{SD}}$ is varied in simulation according to an uncertainty in the mass scale and the mass resolution.  We consider only the impact on selection efficiency from this variation.  The mass is varied by $0.94\%$ to account for the scale, and the resolution on the mass is varied by $20\%$.  These scale factor and mass uncertainties are derived in Ref.~\cite{CMS:2017wyc}.

\item {\cPqb{} tagging scale factor uncertainty (cut-based)}: We apply scale factors to account for the difference in the \cPqb{} jet tagging efficiency between simulation and data \cite{BTV-16-002}. This factor, as well as its uncertainty, depends on the \pt, $\eta$, and hadron flavor of the jet.  This affects the shape of the \htaksml distribution, and is applied by varying the scale factor of \cPqb{} and \cPqc{} jets simultaneously. Light-jet weights are varied separately, resulting in two separate systematic uncertainties.

\item{Extrapolation fit (cut-based)}: The function we use to correct the QCD multijet background prediction from data carries some statistical uncertainty from the fitting procedure.  We assign a corresponding systematic uncertainty equal to the combined uncertainty on the fit parameters.  We generate templates by shifting the fitted function by these uncertainties, and reevaluating the background in each bin.  There is one fit per \PW{} tag category, and therefore two independent nuisance parameters.  These are correlated across \cPqb{} tag categories with equal \PW{} tags.

\item{Normalization systematic (cut-based)}: Sideband regions with 0 \cPqb{}-tagged jets and 1 or 2 \PW{}-tagged jets are used to validate the cut-based analysis method. A small normalization discrepancy is observed after applying the QCD multijet background estimation technique.  Two conservative, independent, log-normal nuisance parameters are therefore included for the QCD multijet background estimation, one applying to the 1\PW{} categories and one applying to the 2\PW{} categories, each with a value of 20\%.  We perform a maximum likelihood fit using only the $0$ \cPqb{} tag sideband categories, and extract scale factors and associated uncertainties for these two parameters.  The extracted scale factors are then applied to the signal regions as listed in Table~\ref{tab:systematics}.
\end{itemize}

\section{Signal discrimination}
\label{sec:disc}

\begin{figure*}
  \begin{center}
    \includegraphics[width=0.45\textwidth]{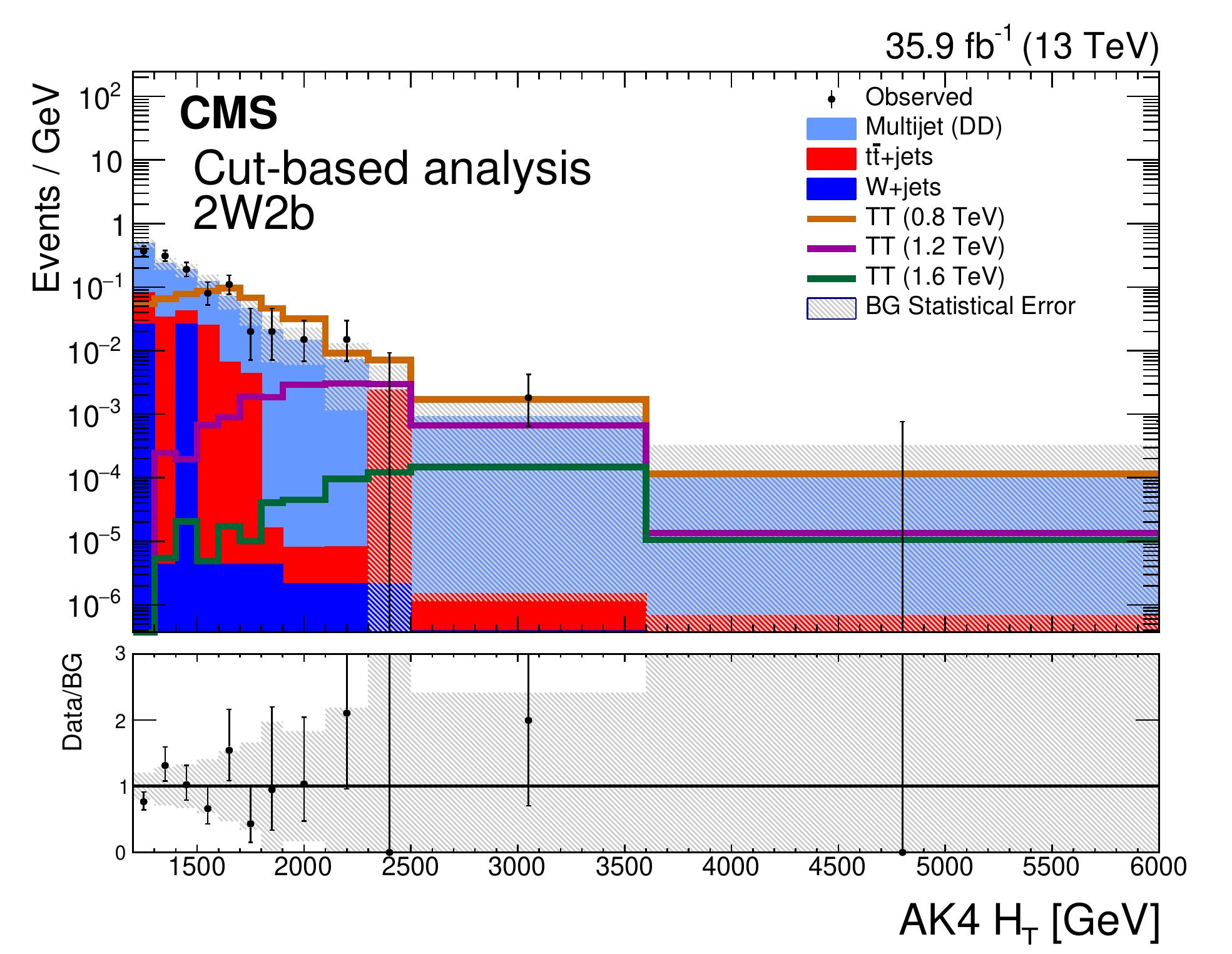}
    \includegraphics[width=0.45\textwidth]{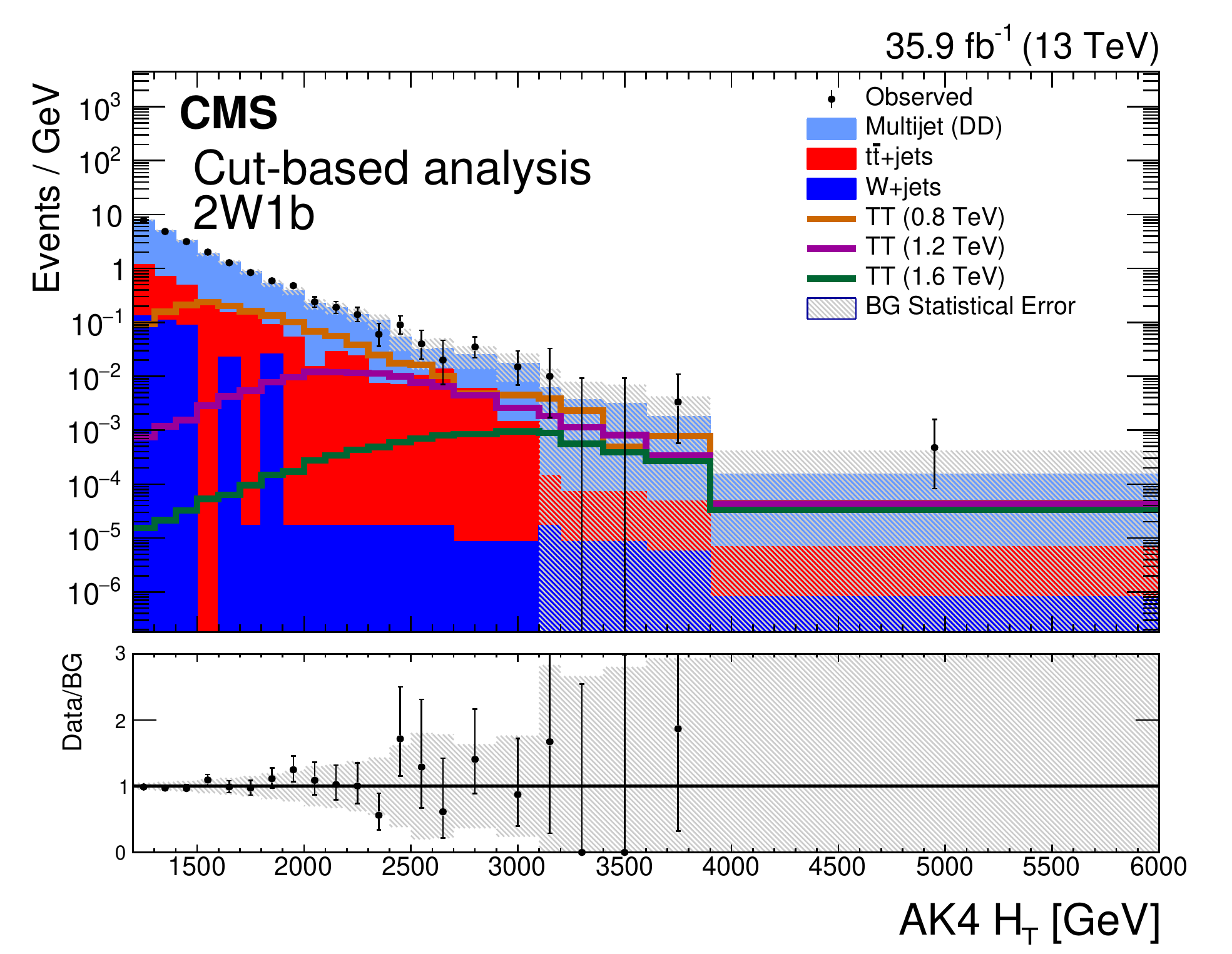}\\
    \includegraphics[width=0.45\textwidth]{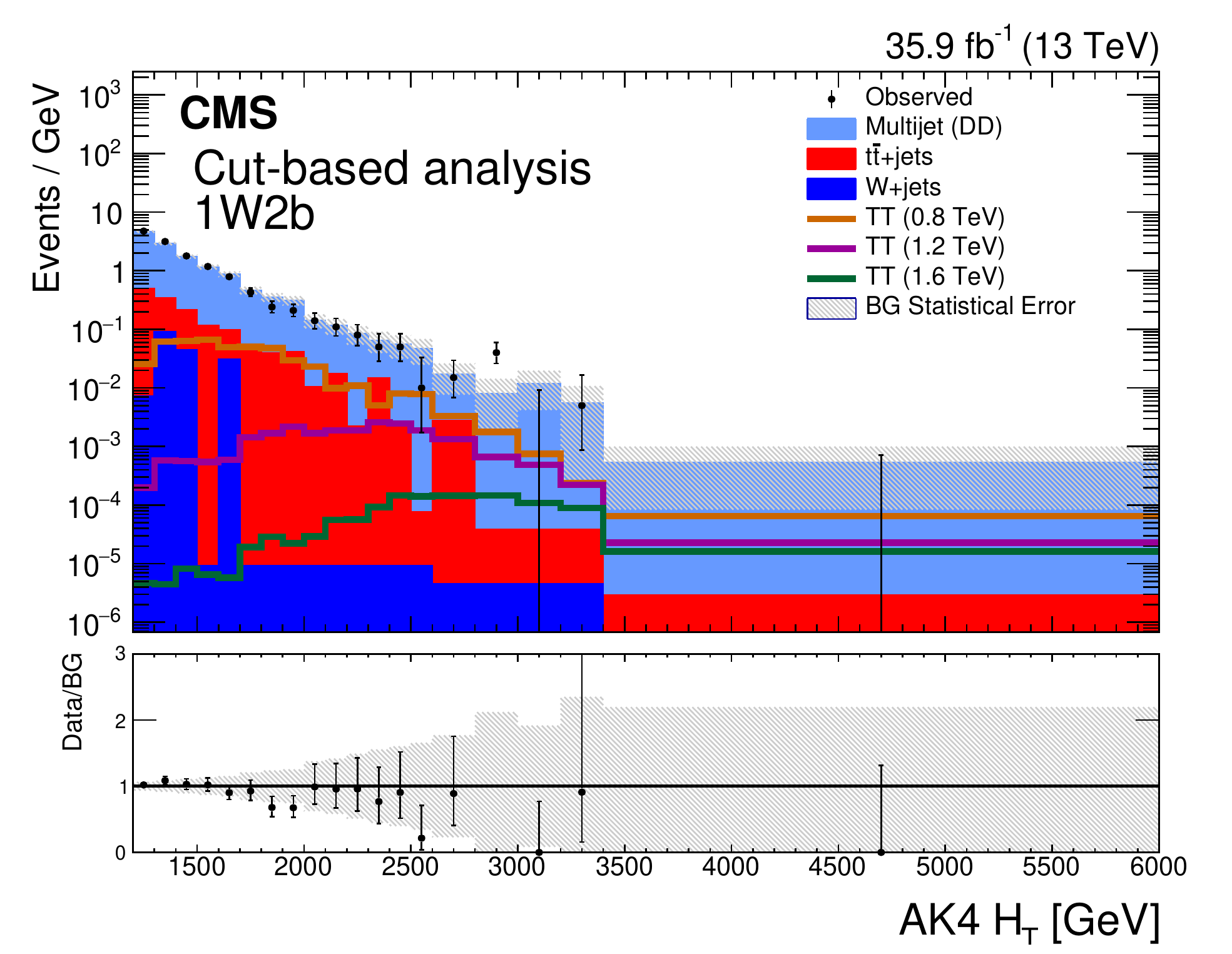}
    \includegraphics[width=0.45\textwidth]{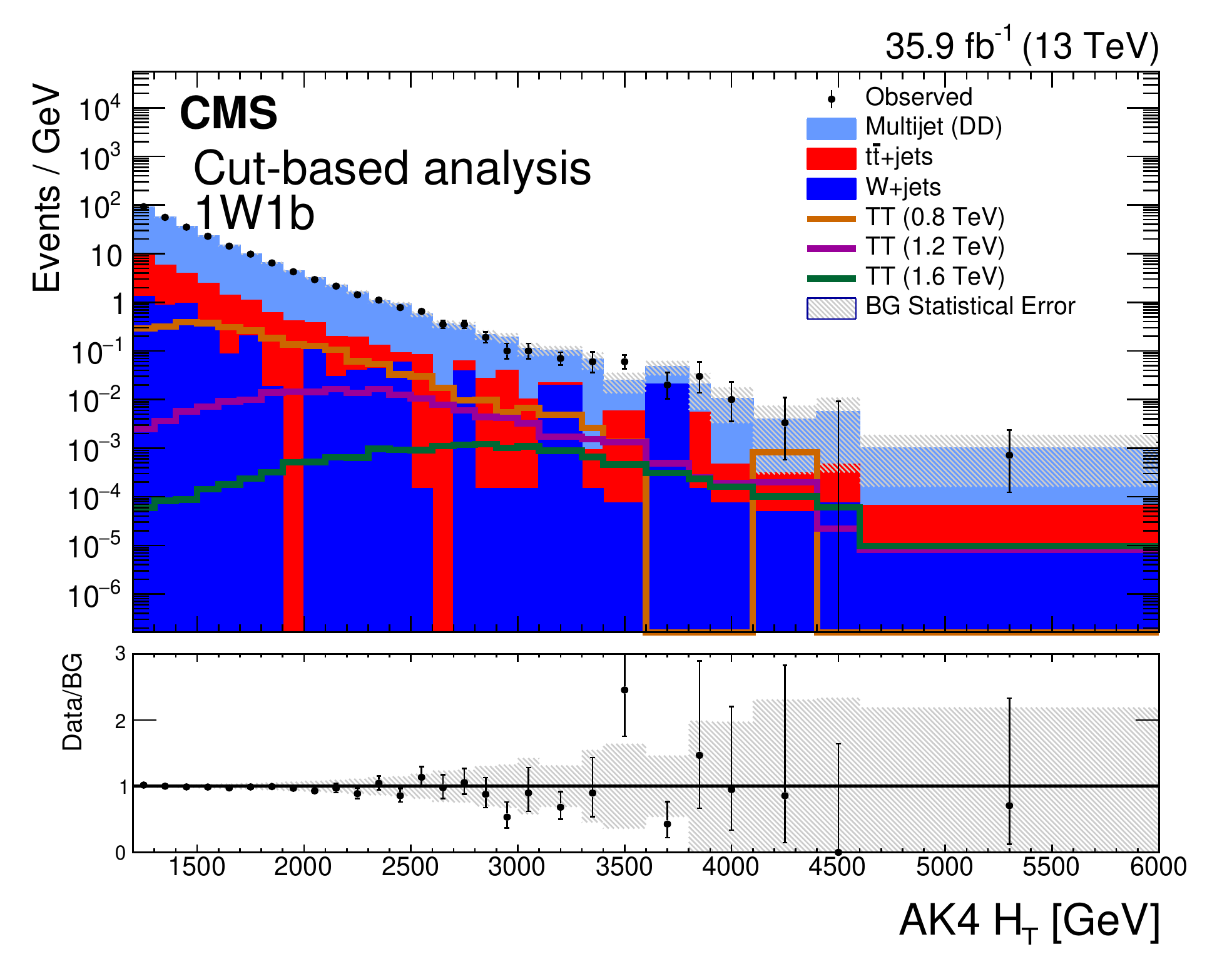}\\
    \caption{The distributions of \htaksml for each of the four signal region categories in the cut-based analysis. The upper row shows channels with $2$ \PW{} tags, and $2$ or $1$ \cPqb{} tags, respectively.  The lower row is for $1$ \PW{} tag. The shaded error band represents the statistical uncertainty in the background. These distributions reflect the nuisance parameters evaluated after a likelihood fit to a background plus signal hypothesis, where the hypothesized signal is a \PQT quark with a mass of 1200\GeV and 100\% branching fraction to \cPqb{}\PW{}. The signal distributions show the expected yield of events assuming the cross section values in Table \ref{signalXS}.  The vertical axis labels denote that bin contents in these distributions have been scaled by their corresponding bin widths, which are the widths used in the fit. The lower panel of each plot shows the ratio of the observed number of events in a bin to the expected number.}
    \label{fig:cutbasedSR}
    \end{center}
\end{figure*}

In this section, we present the distributions used to test for the presence of a signal.  For the cut-based analysis, there are 4 independent categories: 1 \PW{} tag with either 1 or 2 \cPqb{}-tagged jets, and 2 \PW{} tags with either 1 or 2 \cPqb{}-tagged jets.  In each category, the \htaksml distribution is used for signal discrimination.  Figure \ref{fig:cutbasedSR} shows the \htaksml distributions for each of the 4 signal region categories. The amount of signal that falls into these categories depends on the hypothesized mass and decay fraction; for a \cPqb{}\PW{}\cPqb{}\PW{} decay, the acceptance ranges from 6.1 to 7.5\%. For a \cPqt{}\PZ{}\cPqt{}\PZ{} decay, the range is 3.8 to 7.5\%, and for \cPqt{}\PH{}\cPqt{}\PH{} it is 3.6 to 6.9\%.

\begin{figure*}
\begin{center}
\includegraphics[width=1.0\textwidth]{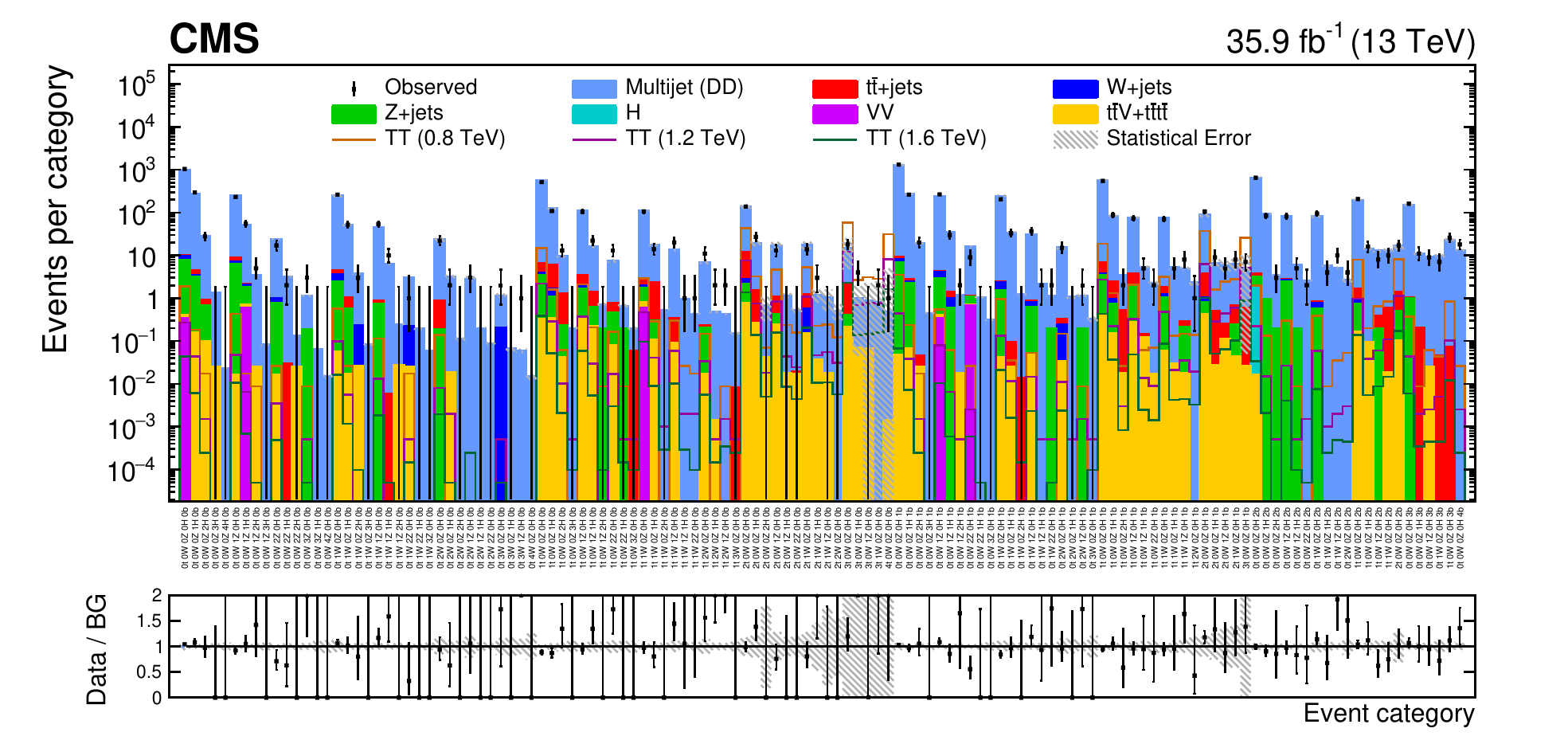}
\caption{A summary of the 126 signal region categories used in the NN analysis.  This figure shows the expected yields in each category, while the signal discrimination is performed with the \htakbig distributions from each of the categories.  The bottom panel shows the ratio of observed data to total background in each category, with Poisson error bars where applicable, along with the total background uncertainty shown for each category by the gray band.}
\label{fig:NNSR}
\end{center}
\end{figure*}

\begin{figure*}
  \begin{center}
    \includegraphics[width=0.45\textwidth]{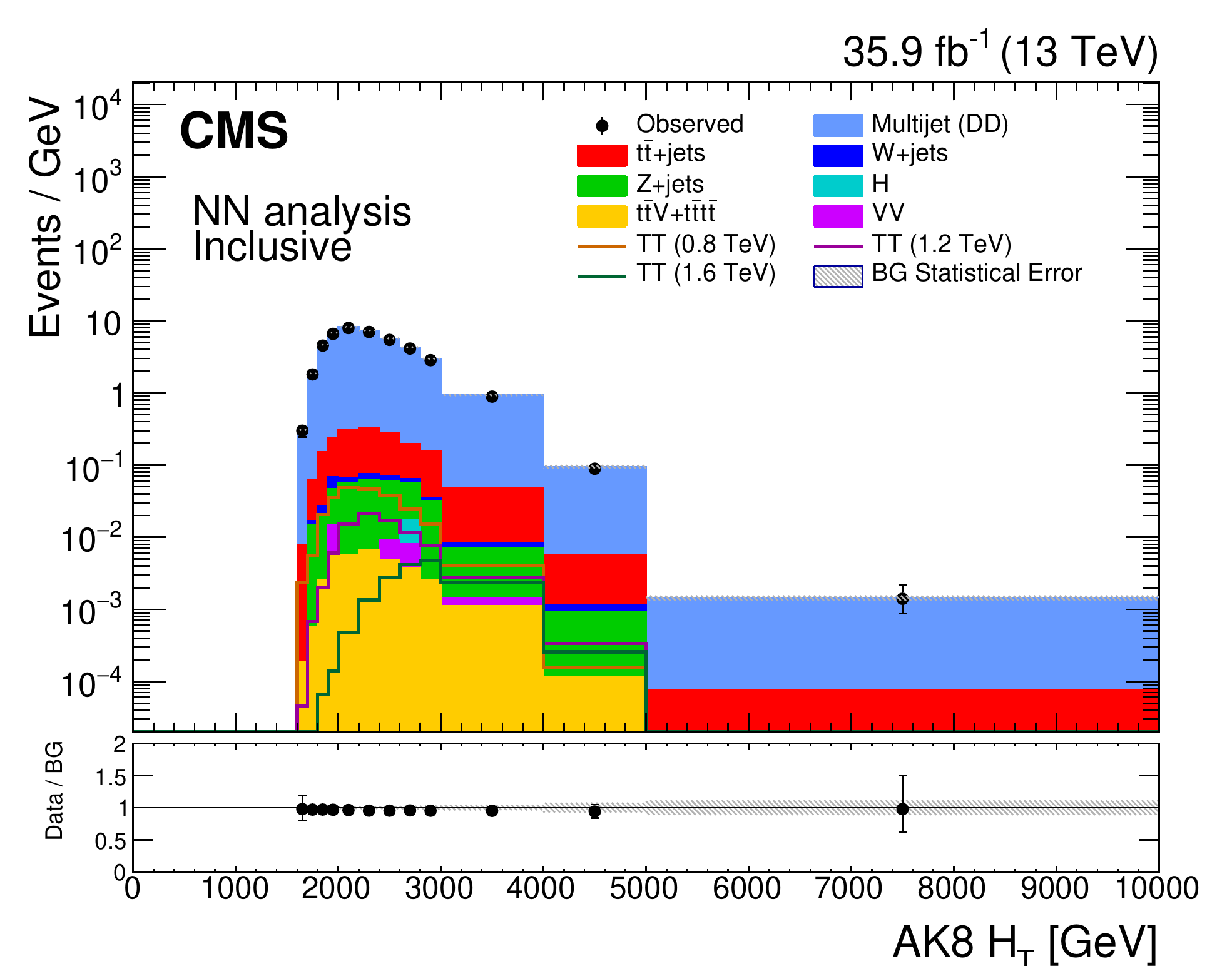}
    \includegraphics[width=0.45\textwidth]{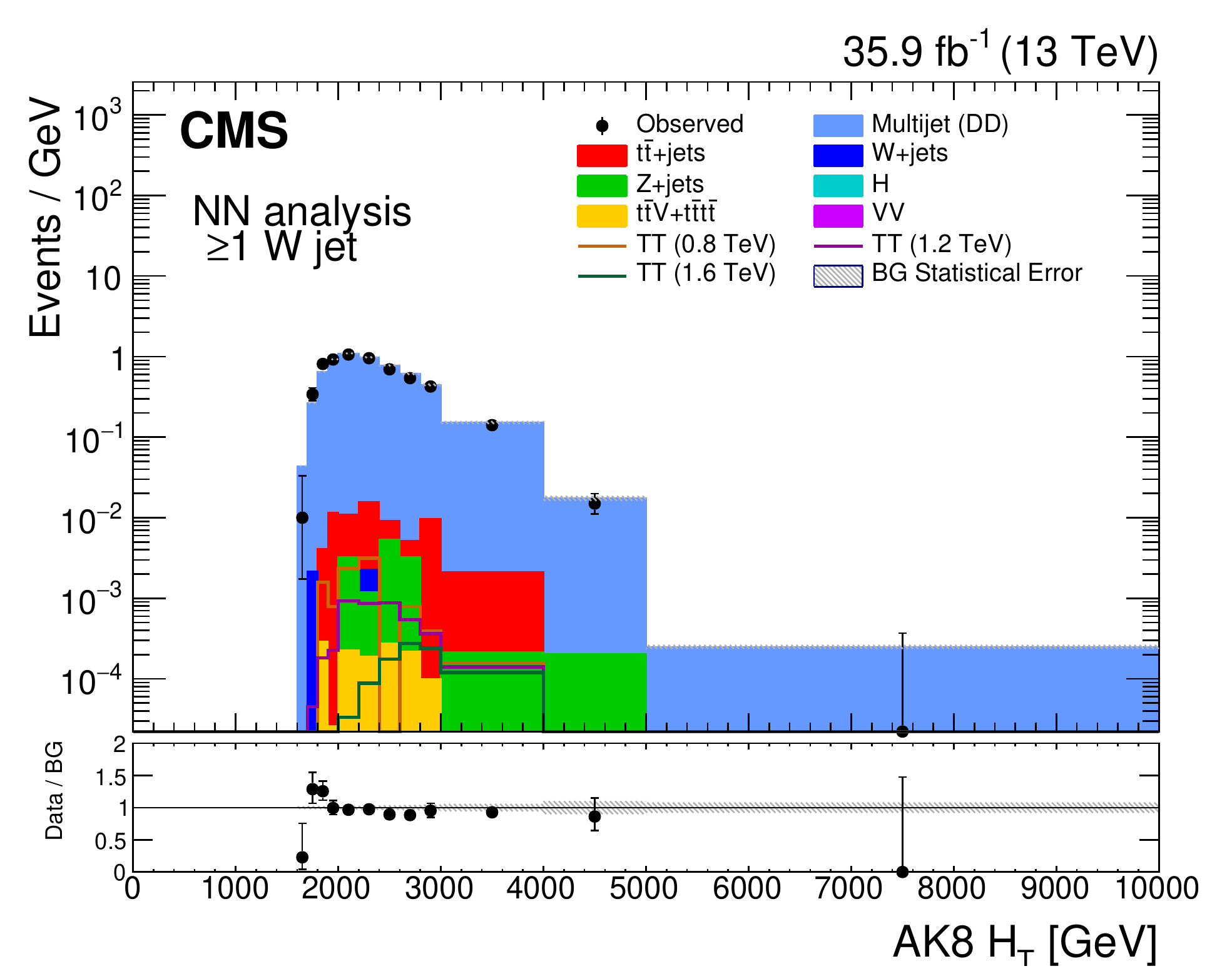}\\
    \includegraphics[width=0.45\textwidth]{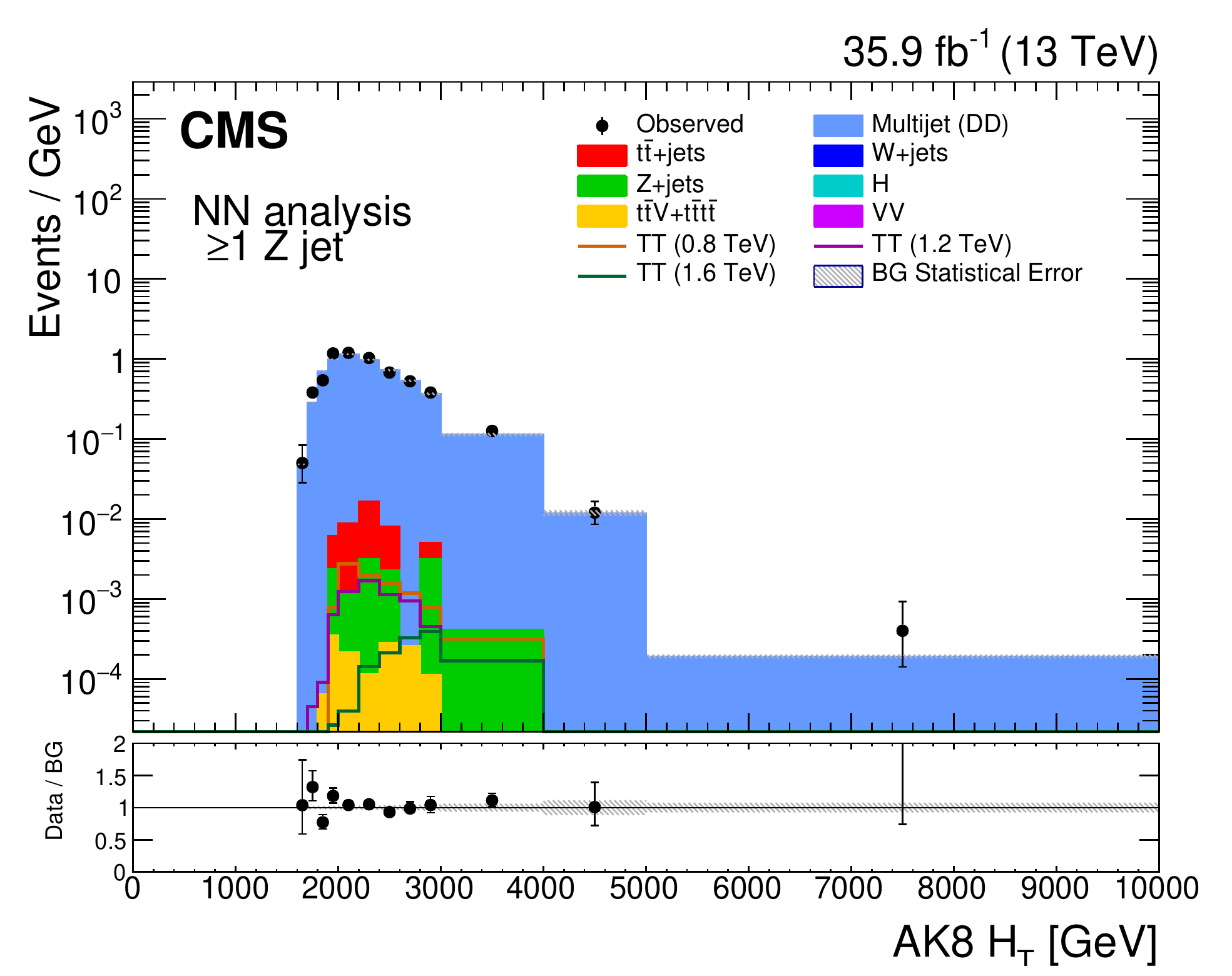}
    \includegraphics[width=0.45\textwidth]{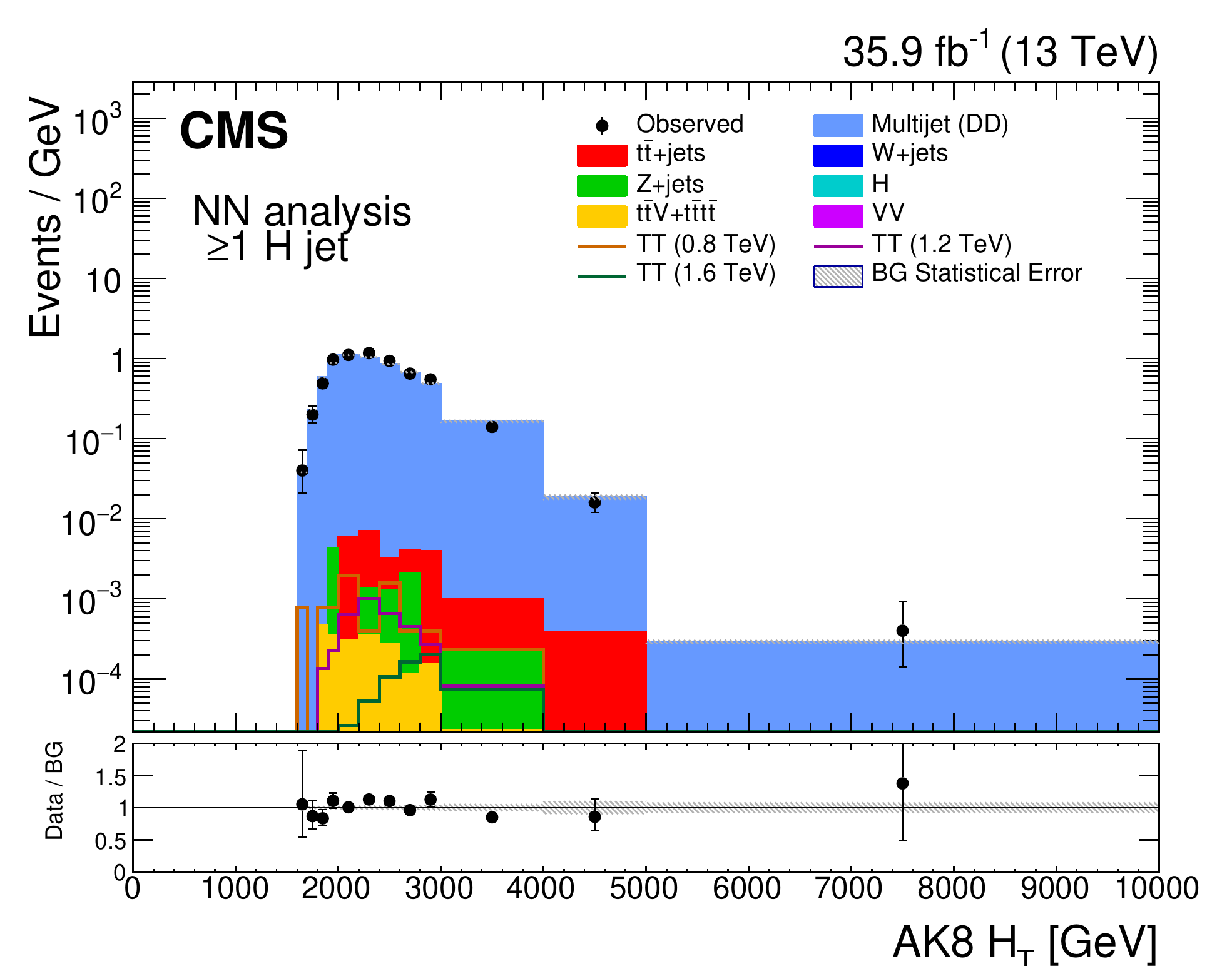}\\
    \includegraphics[width=0.45\textwidth]{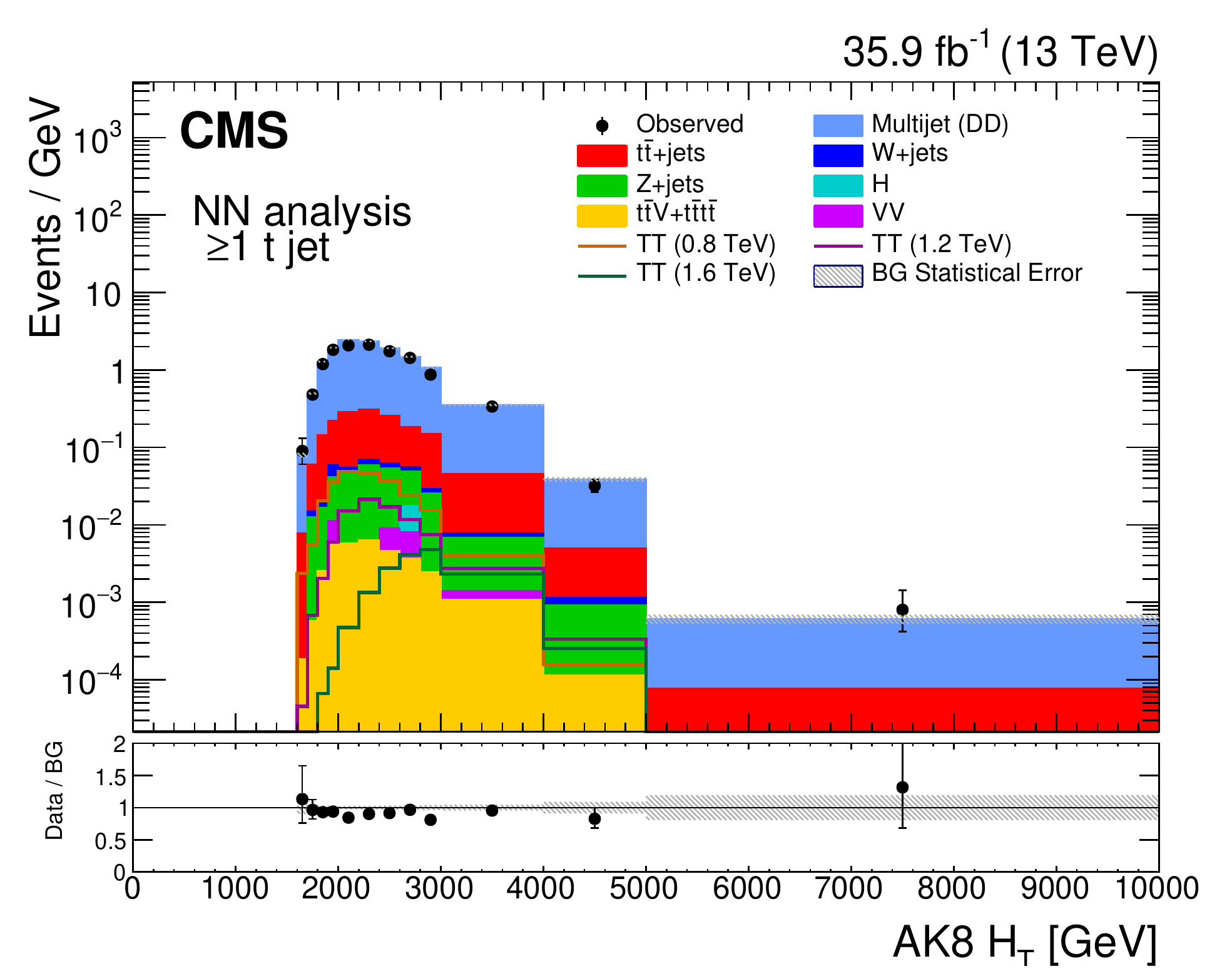}
    \includegraphics[width=0.45\textwidth]{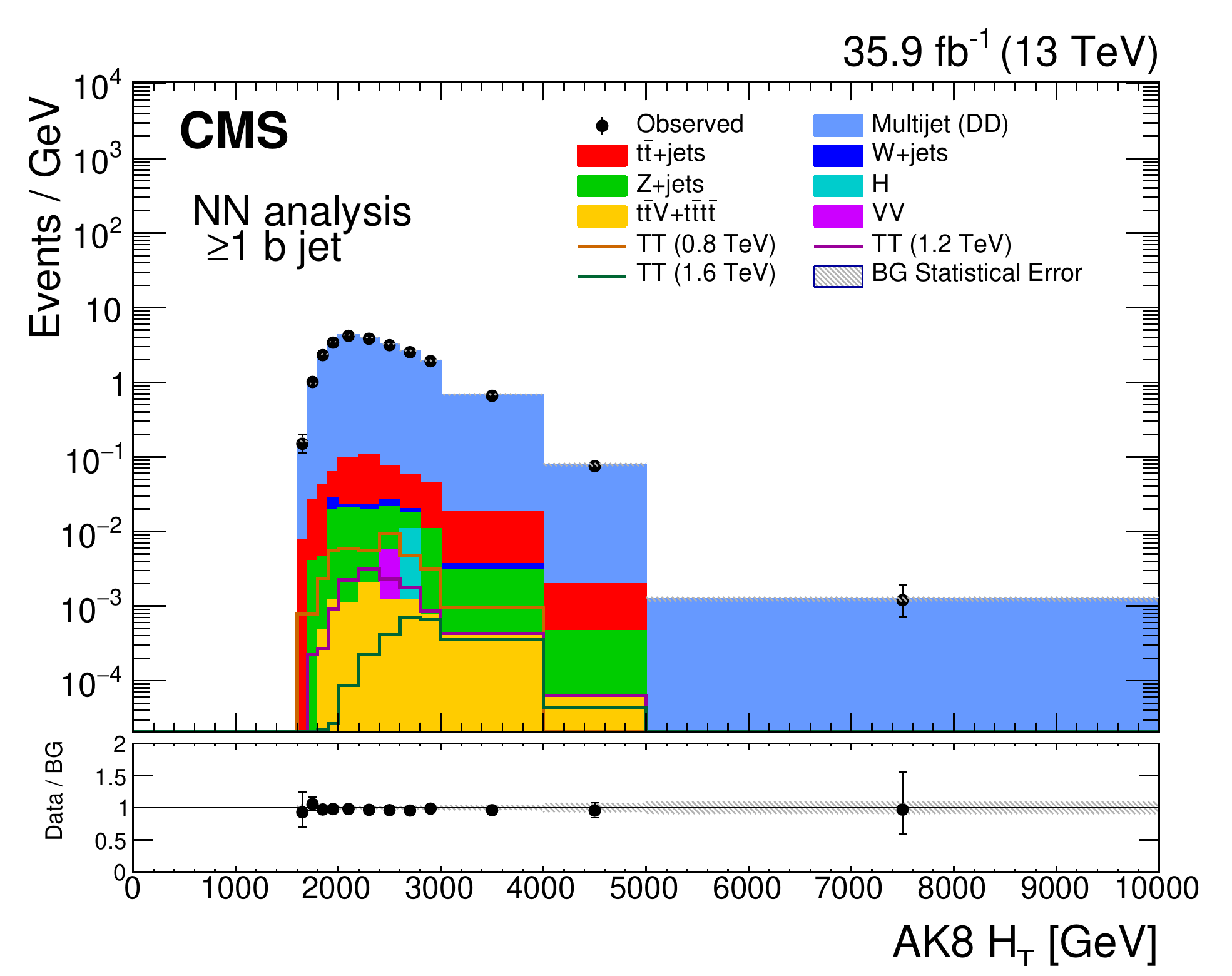}\\
    \caption{Distributions of \htakbig for all events entering the 126 signal regions of the NN analysis (upper left), as well as for only categories containing at least one candidate of each of the particle types identified by the BEST algorithm: $\geq$1 \PW{} jet (upper right), $\geq$1 \PZ{} jet (middle left), $\geq$1 \PH{} jet (middle right), $\geq$1 \cPqt{} jet (lower left), and $\geq$1 \cPqb{} jet (lower right).  The plots shown here are not mutually exclusive, as a particular signal region may satisfy several of the criteria for the individual summary categories.  The vertical axis labels denote that bin contents in these distributions have been scaled by their corresponding bin widths.  The lower panel of each plot shows the ratio of the observed number of events in a bin to the expected number.}
    \label{fig:NNht}
    \end{center}
\end{figure*}

For the NN analysis, there are 126 independent signal region categories, corresponding to all the possible combinations of BEST label multiplicities for 4 AK8 jet events.  Between $0.3\%{}$ and $15\%{}$ of signal events with a \cPqt{}\PZ{}\cPqt{}\PZ{} decay pass the kinematic requirements to be placed into these signal regions, depending on the VLQ mass. For a \cPqb{}\PW{}\cPqb{}\PW{} decay, the range is 0.47\% to 16\%, and for \cPqt{}\PH{}\cPqt{}\PH{} it is 0.33\% to 21\%. Figure \ref{fig:NNSR} shows a visualization of the expected and observed yields in each of the 126 categories.  For further signal discrimination, the analysis results use the \htakbig distribution in each of the signal region categories.  Figure \ref{fig:NNht} shows the \htakbig distributions for combined categories including at least one \PW{}, \PZ{}, \PH{}, \cPqt{}, or \cPqb{} candidate, as well as the inclusive distribution summing all 126 signal regions.  The individual distributions shown in Fig. \ref{fig:NNht} are not independent, as a particular category may satisfy the criteria for several distributions.

\section{Statistical analysis}
\label{sec:stats}

Independent statistical procedures are performed for the cut-based and NN analyses, using the same methodology.  No explicit combination of the two analyses is presented here, as they are performed on many of the same events.  The \textsc {Theta} software package \cite{theta} is used to perform a Bayesian shape-based analysis using the distributions from the signal region categories.  Each bin of the distributions is combined statistically in a likelihood, where contributions from systematic and statistical uncertainties are added through nuisance parameters in the likelihood function.  Each of the rate nuisance parameters is implemented with a log-normal prior distribution, while the shape-based nuisance parameters utilize Gaussian prior distributions.

In the cut-based analysis, all four signal regions are fit simultaneously.  Most systematic uncertainties are fit simultaneously across the four categories, with the exception of the extrapolation fit and normalization uncertainties.  For these parameters, there are independent uncertainties for the two \PW tag multiplicities. The ratio of events in the control regions to signal regions is fixed when calculating the multijet background component, and is not a parameter considered in the fit.

For the NN analysis, nuisance parameters for the BEST classification efficiency scale factors are allowed to fluctuate unconstrained, allowing a simultaneous measurement of scale factor values. A uniform prior distribution is assumed for the signal normalization.  Additionally, due to the limited numbers of simulated events, we follow the ``Barlow--Beeston lite'' method \cite{Barlow} and assign an additional nuisance parameter to each bin of background components relying on simulated events. Prior to the statistical analysis, the discriminating distributions are rebinned from a width of 100\GeV{}, to reduce the statistical uncertainty on the total background in the tails of the distributions to below $30\%$, as they can suffer from the effects of having limited events passing all signal criteria.  The likelihood function is used to extract Bayesian upper limits on the cross section for pair production of \PQT or \PQB quarks at $95\%$ confidence level (\CL{}).  Additionally, samples of pseudodata are formed by sampling the expected backgrounds after varying the uncertainties within their prior distributions.  For each pseudodata sample, the statistical analysis is performed to extract a range of upper limit outcomes.  The median of these outcomes is the expected limit, and the range of outcomes within one or two standard deviations of the median is also shown for comparison.

\section{Results}
\label{sec:results}

\begin{figure*}
\begin{center}
\includegraphics[width=0.4\textwidth]{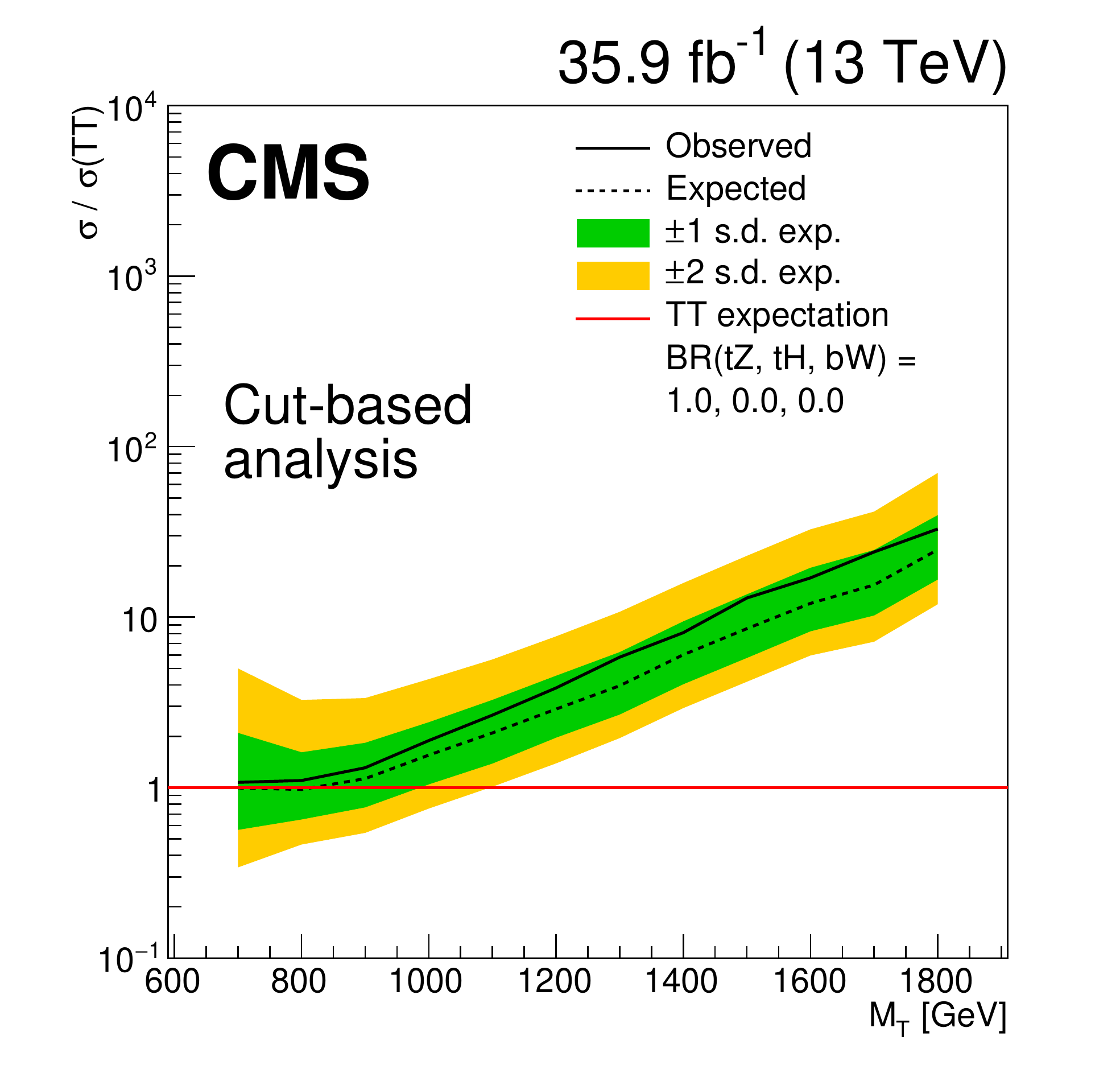}
\includegraphics[width=0.4\textwidth]{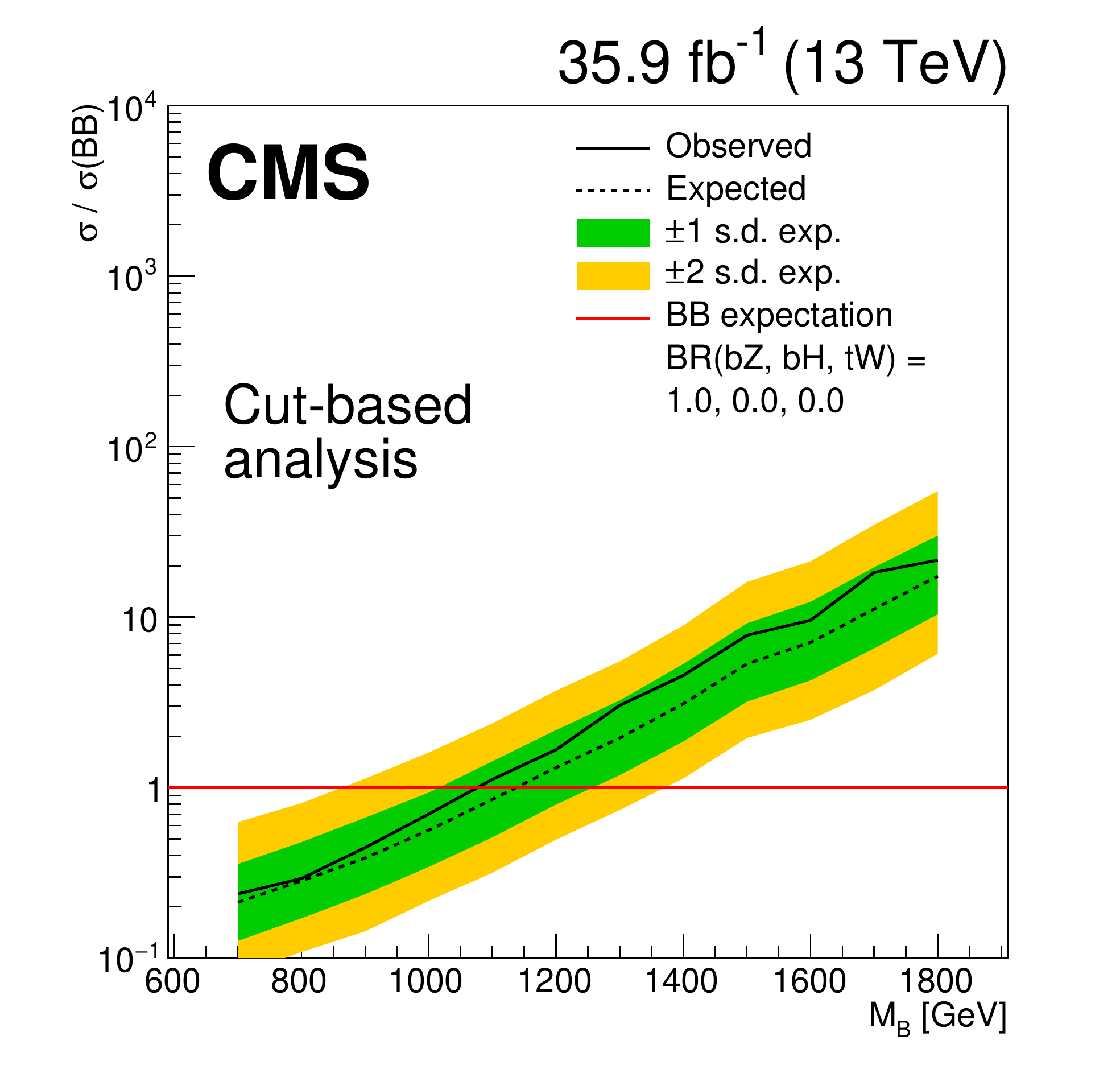}\\
\includegraphics[width=0.4\textwidth]{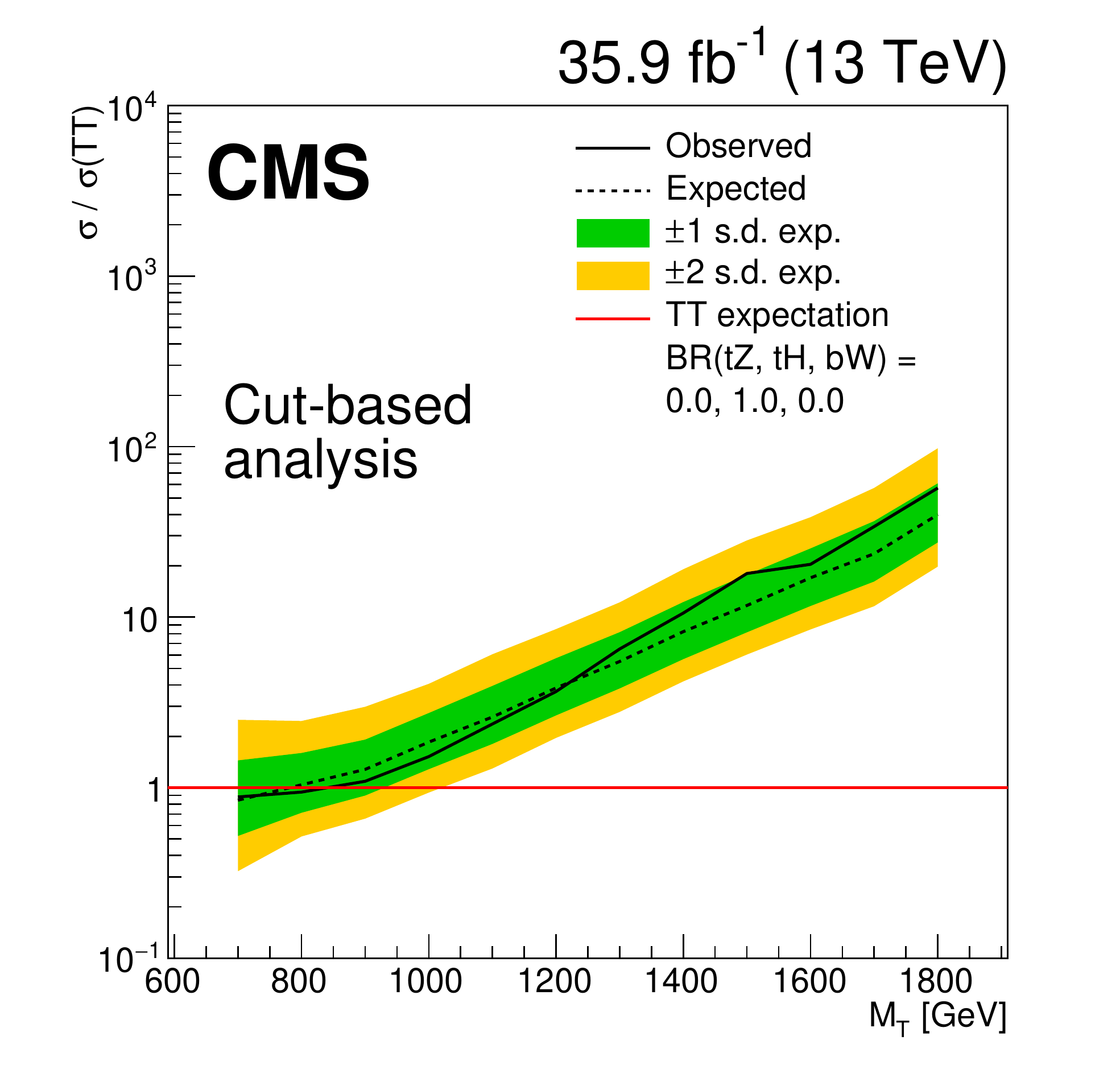}
\includegraphics[width=0.4\textwidth]{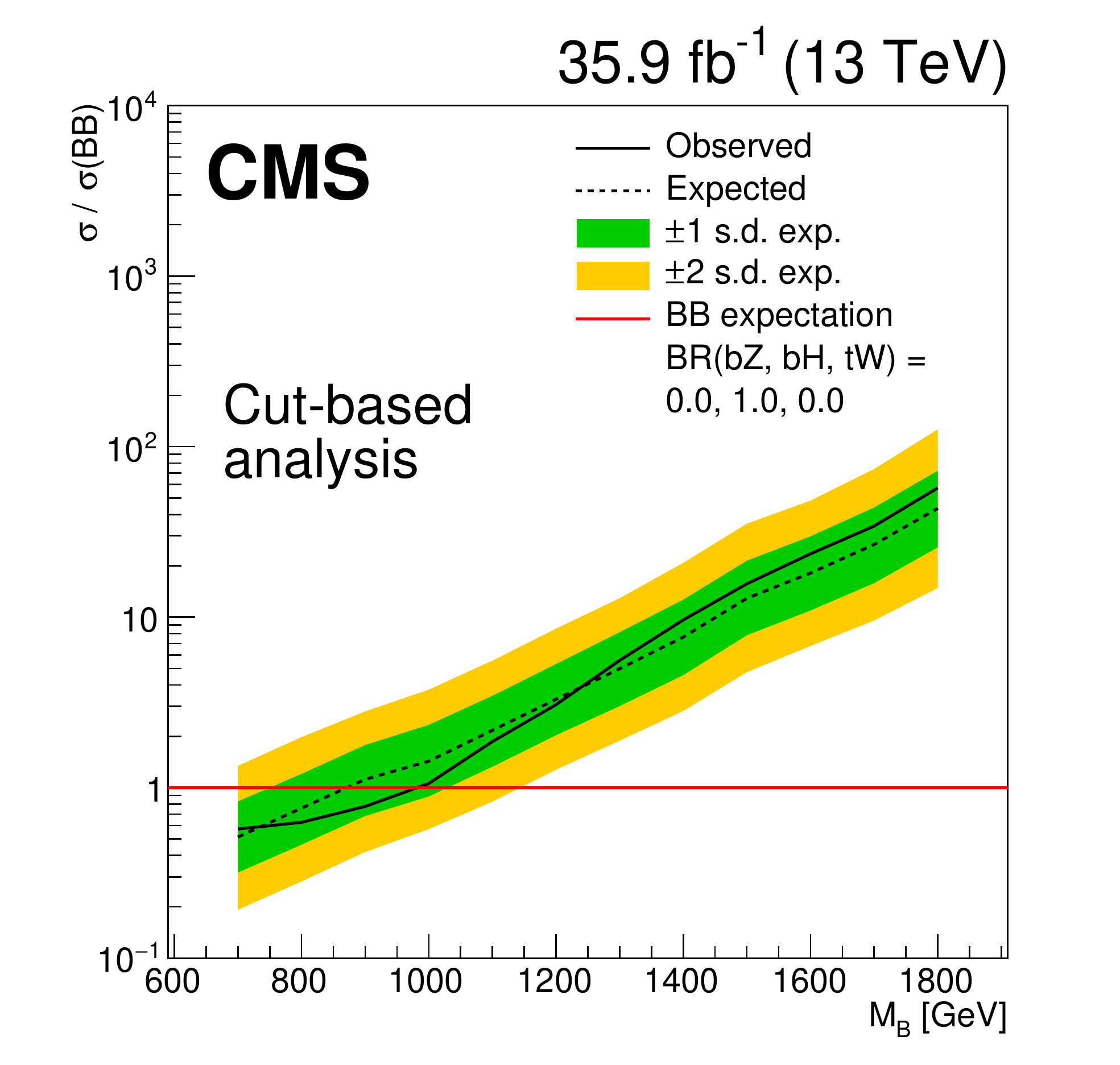}\\
\includegraphics[width=0.4\textwidth]{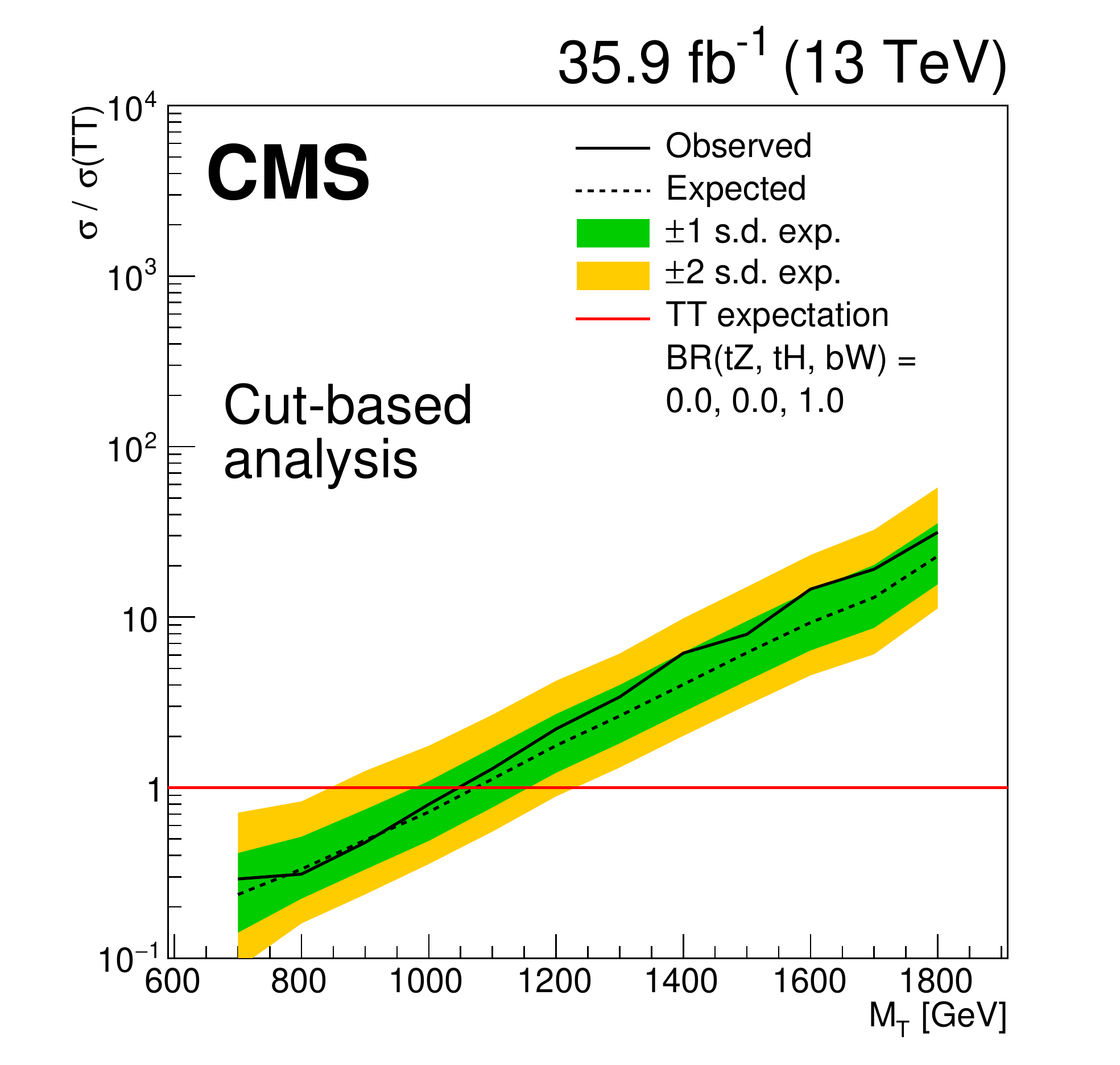}
\includegraphics[width=0.4\textwidth]{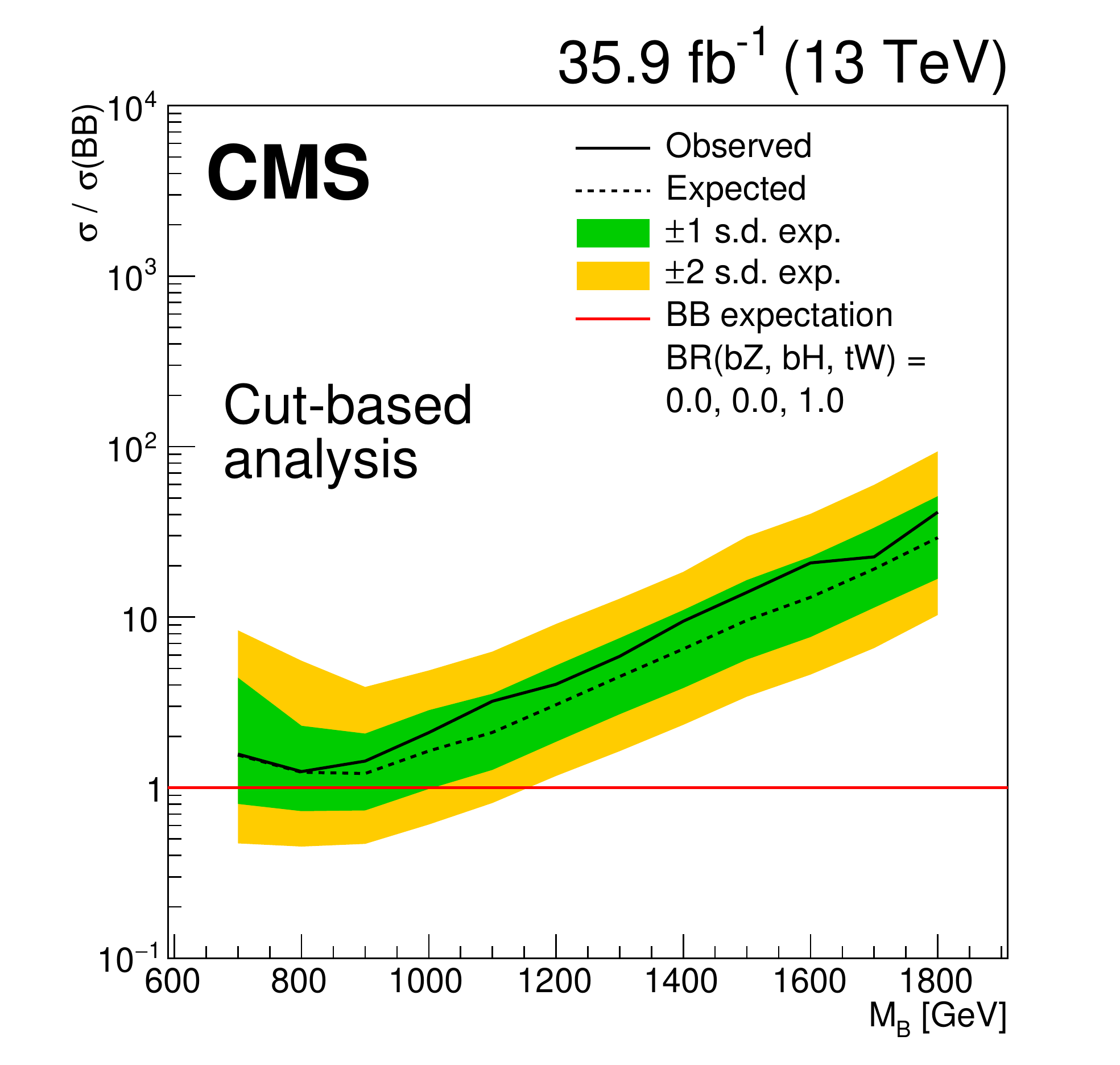}
\caption{Limits at $95\%{}$ confidence level on the ratio of the cross section to the theoretical cross section for pair production \PQT quarks (left) and \PQB quarks (right) in the cut-based analysis, with decays solely to \cPqt{}\PZ{}/\cPqb{}\PZ{} (upper), \cPqt{}\PH{}/\cPqb{}\PH{} (middle), and  \cPqb{}\PW{}/\cPqt{}\PW{} (lower).  The solid black line shows the observed limit, while the dashed black line shows the median of the distribution of limits expected under the background-only hypothesis.  The inner (green) band and the outer (yellow) band indicate the regions containing 68 and 95\%, respectively, of the distribution of limits expected under the background-only hypothesis.}
\label{fig:xslimit_cutbased}
\end{center}
\end{figure*}

\begin{figure*}
\begin{center}
\includegraphics[width=0.4\textwidth]{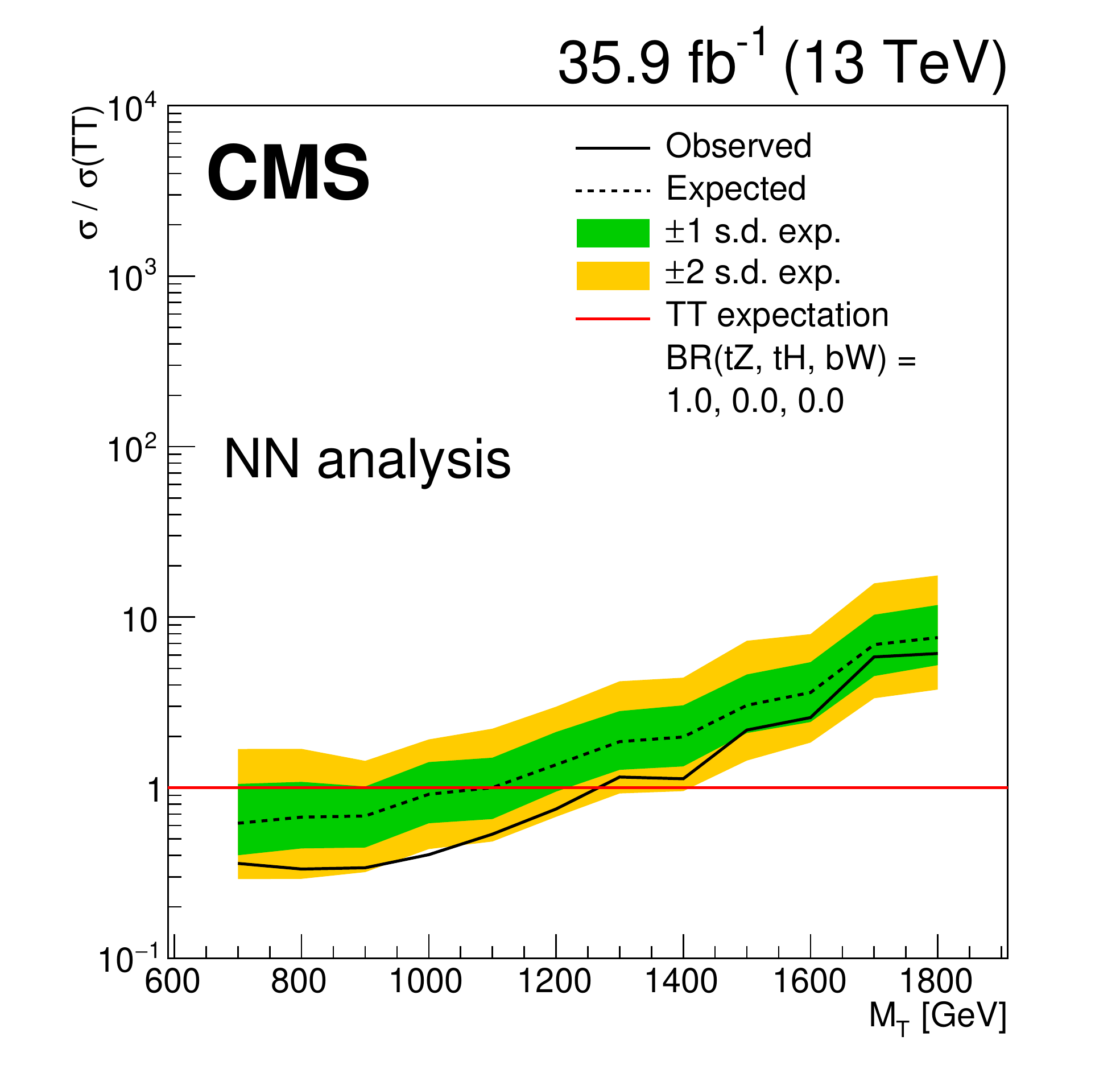}
\includegraphics[width=0.4\textwidth]{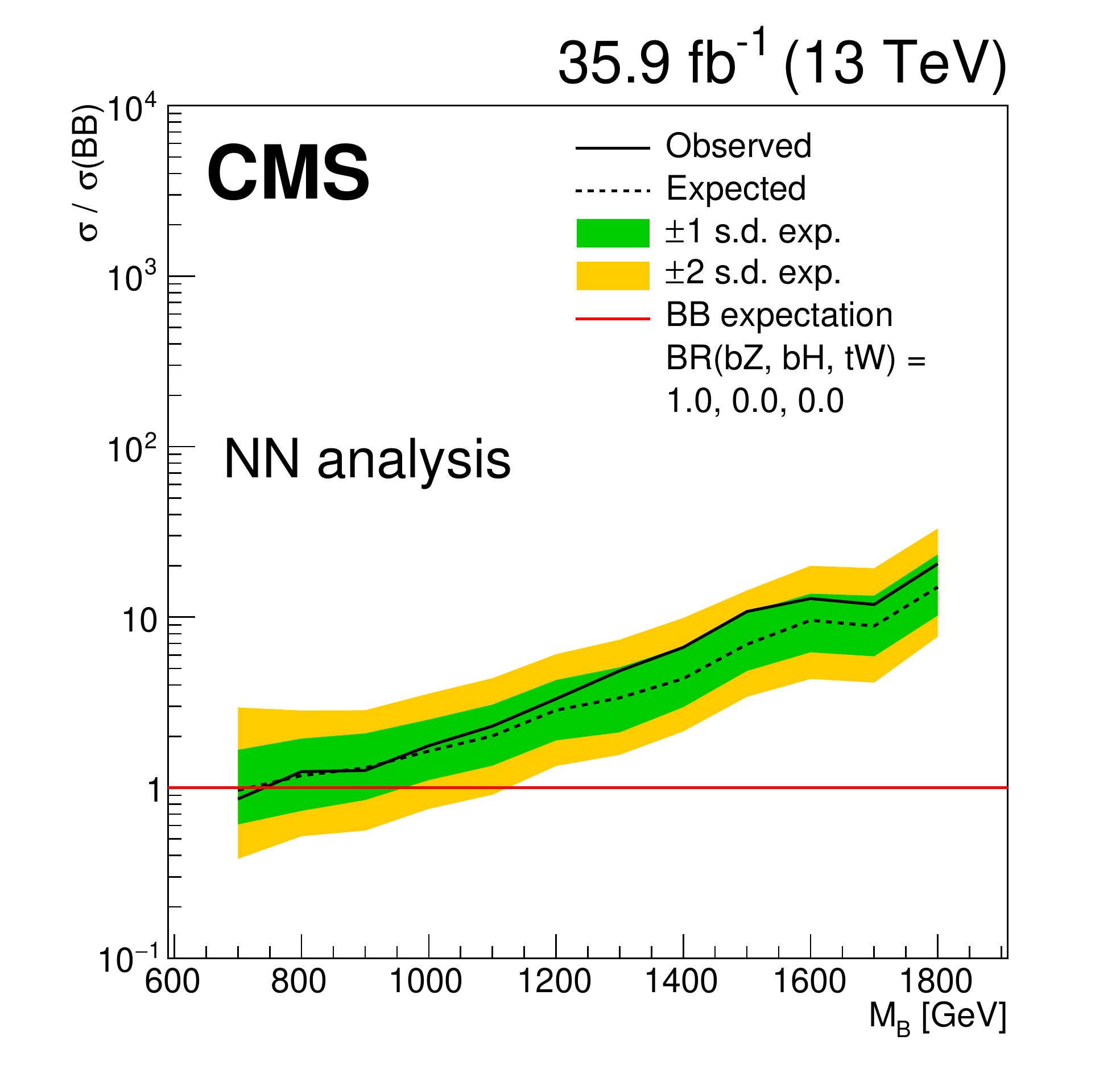}\\
\includegraphics[width=0.4\textwidth]{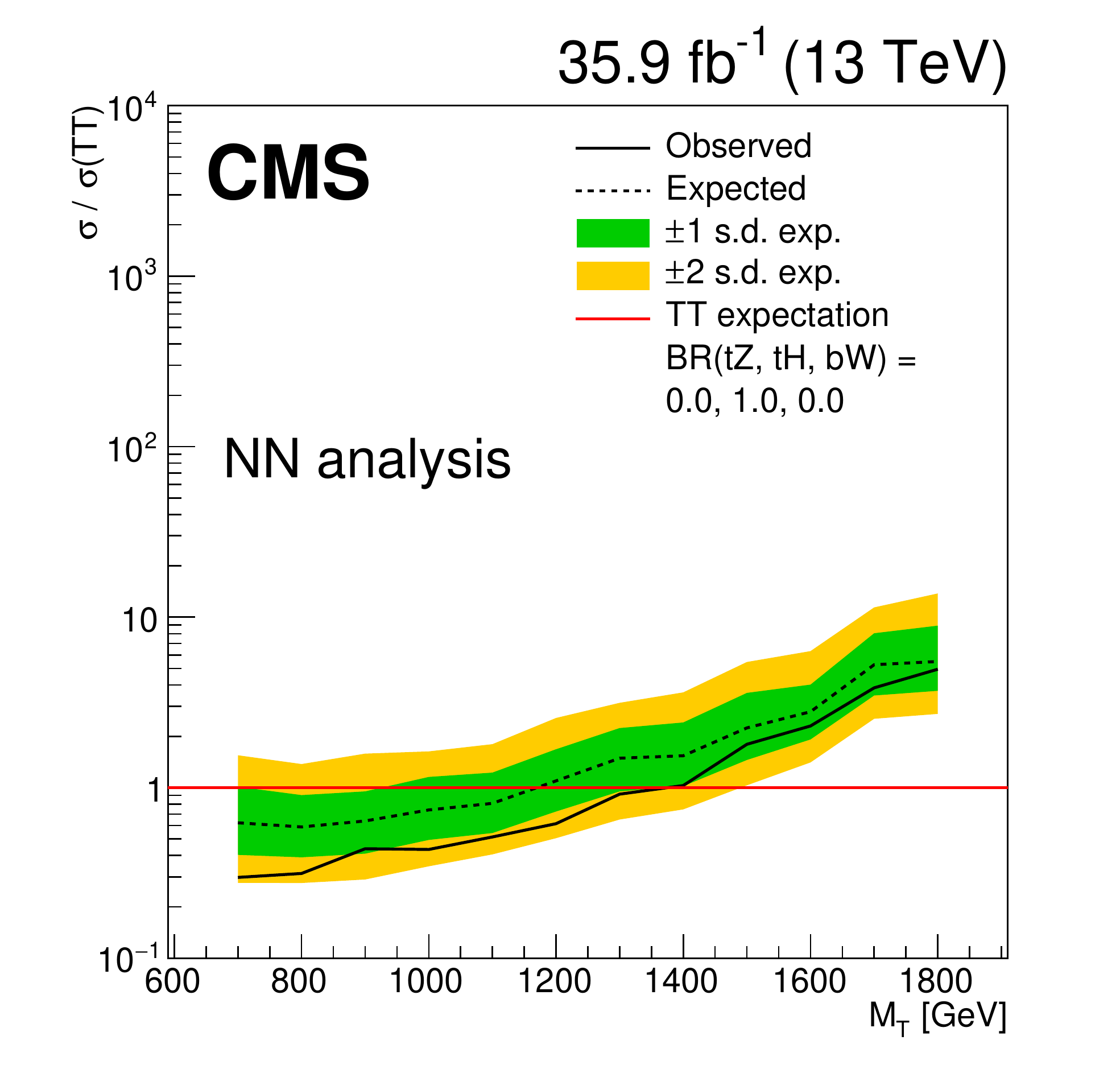}
\includegraphics[width=0.4\textwidth]{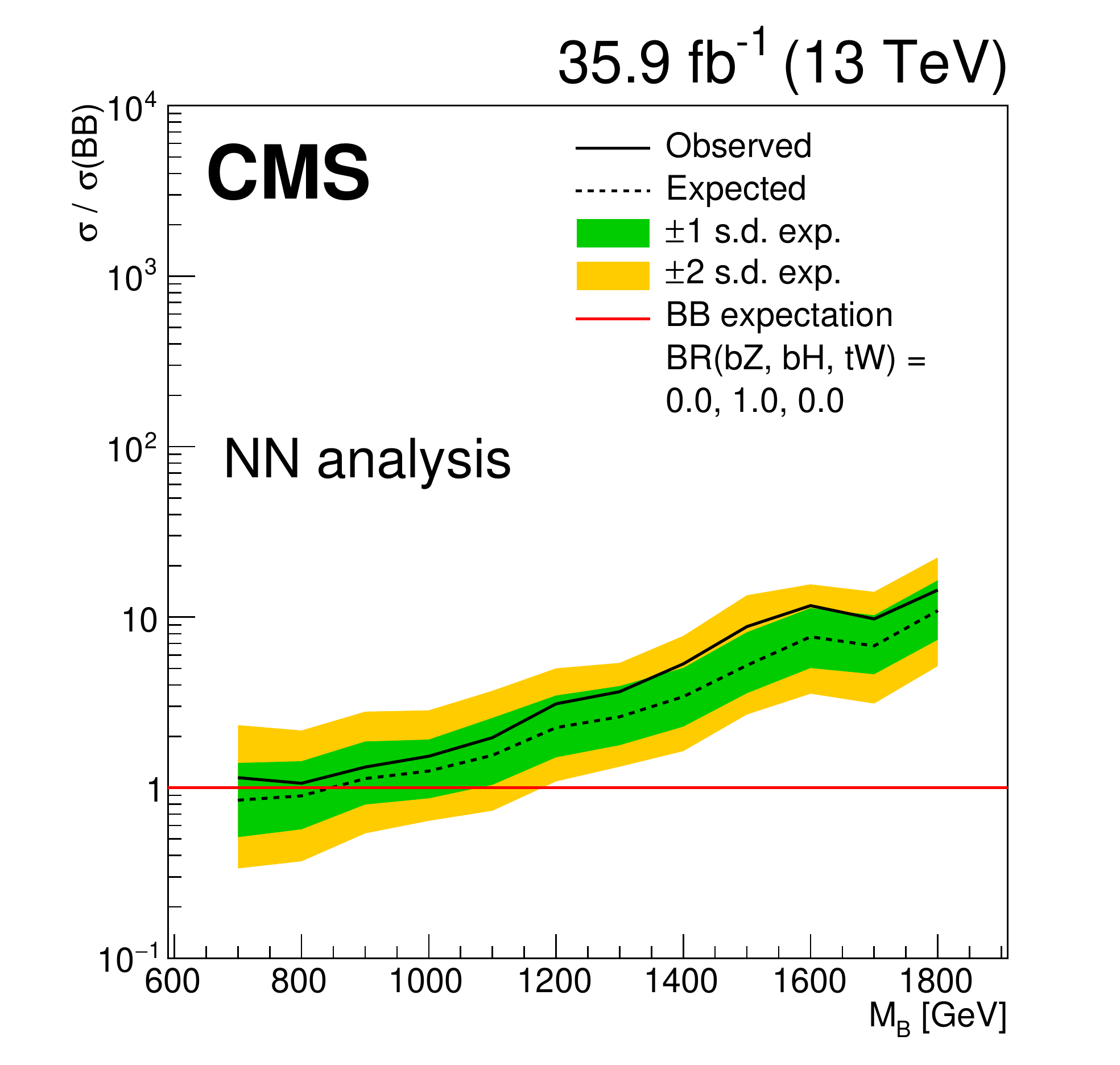}\\
\includegraphics[width=0.4\textwidth]{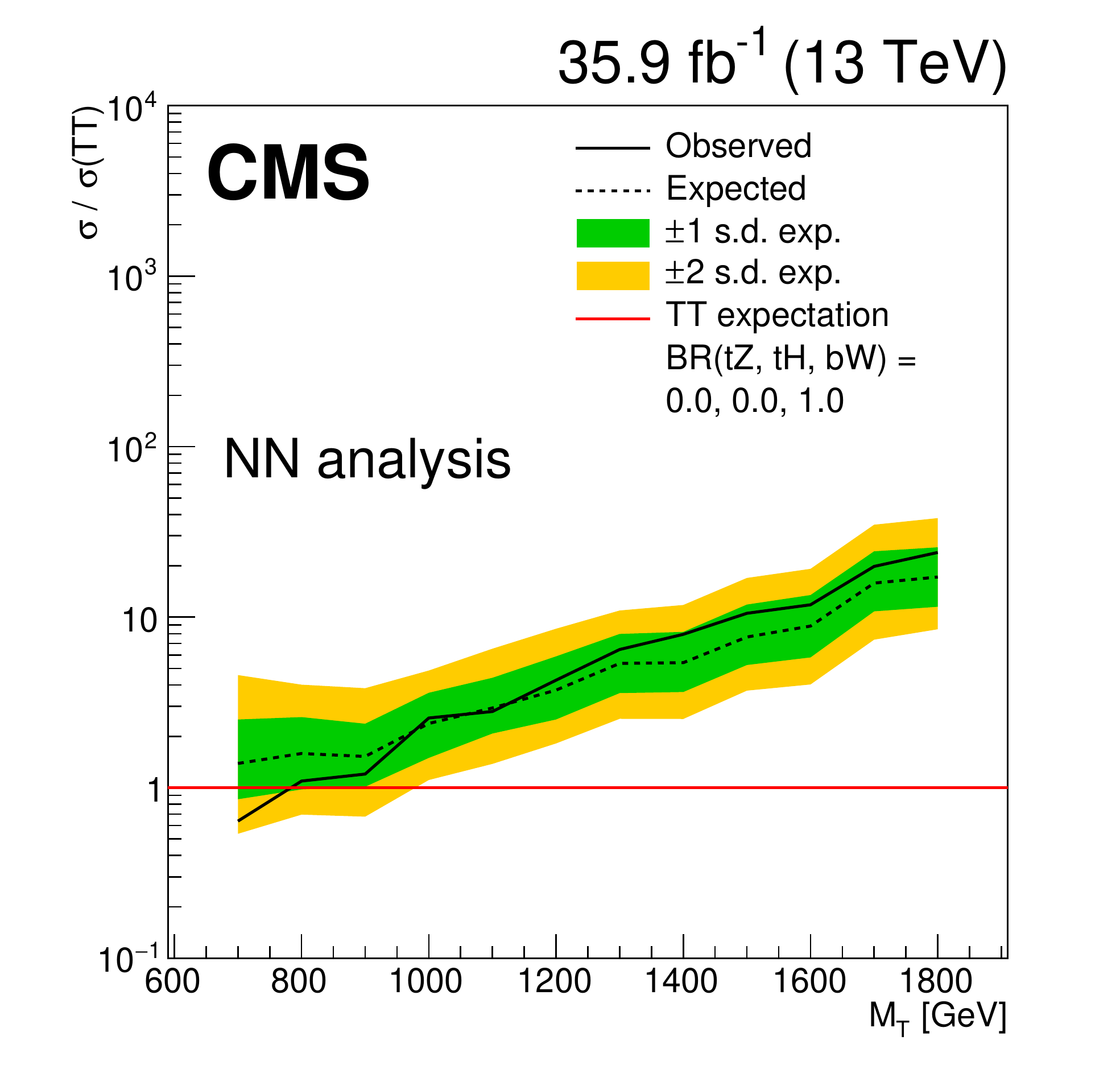}
\includegraphics[width=0.4\textwidth]{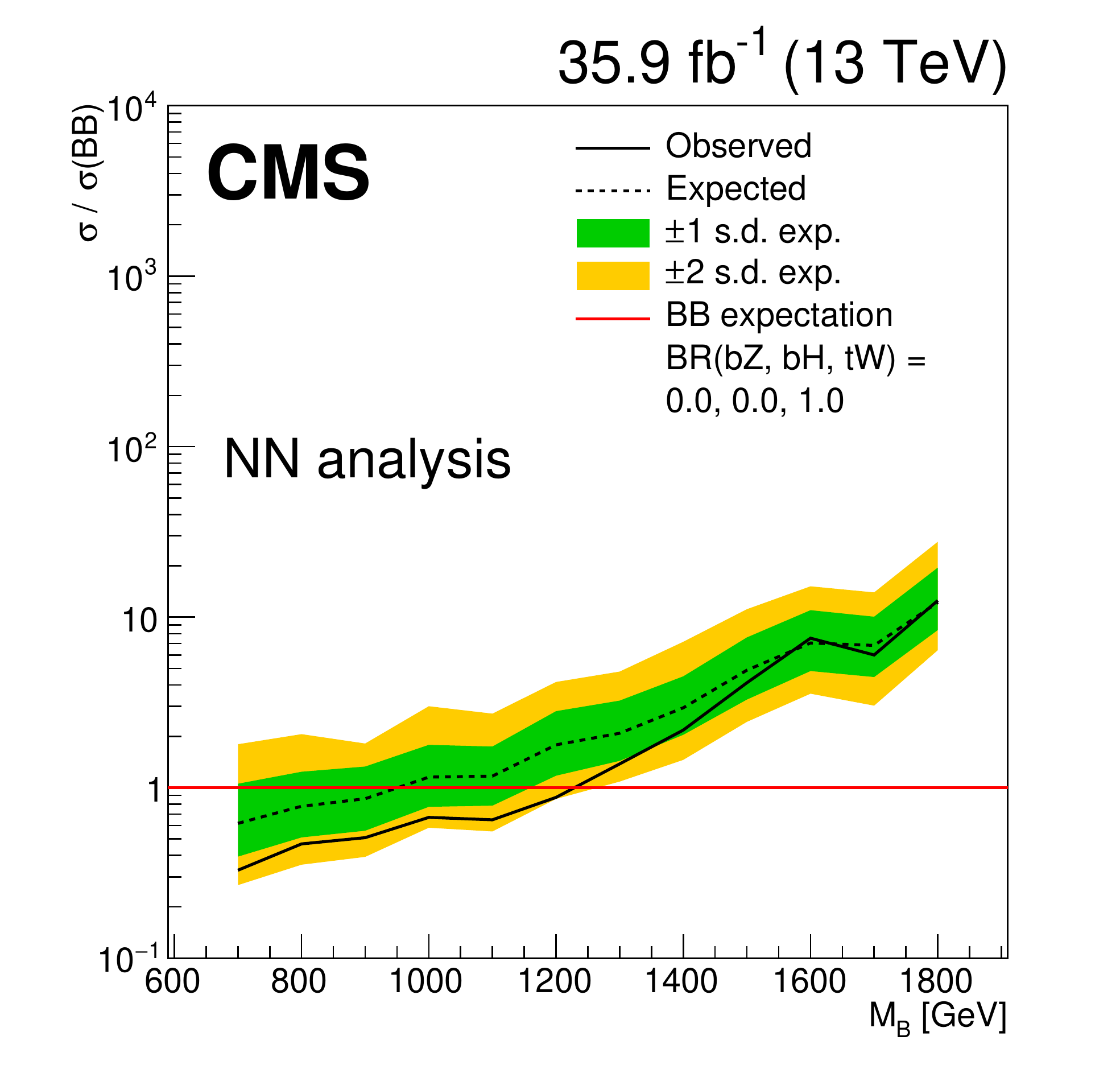}
\caption{Limits at $95\%{}$ confidence level on the ratio of the cross section to the theoretical cross section for pair production \PQT quarks (left) and \PQB quarks (right) in the NN analysis, with decays solely to \cPqt{}\PZ{}/\cPqb{}\PZ{} (upper), \cPqt{}\PH{}/\cPqb{}\PH{} (middle), and \cPqb{}\PW{}/\cPqt{}\PW{} (lower).  The solid black line shows the observed limit, while the dashed black line shows the median of the distribution of limits expected under the background-only hypothesis.  The inner (green) band and the outer (yellow) band indicate the regions containing 68 and 95\%, respectively, of the distribution of limits expected under the background-only hypothesis.}
\label{fig:xslimit_NN}
\end{center}
\end{figure*}

We observe no statistically significant excess over the expected background. The expected and observed limits on the cross section for pair production of \PQT and \PQB quarks are shown in the case of branching fractions of one for the individual decay modes in Figs. \ref{fig:xslimit_cutbased} and \ref{fig:xslimit_NN}, for the cut-based and NN analyses, respectively.  Because the cut-based analysis is optimized for the \cPqb{}\PW{} decay mode and includes selections targeting boosted \PW jets, it lacks sensitivity to the other decay modes.  The NN analysis does not target a specific decay mode, but shows the best sensitivity to \PQT quark decays to \cPqt{}\PZ{} and \cPqt{}\PH{}, or \PQB quark decays to \cPqt{}\PW{}. It has lower sensitivity in the \cPqb{}\PW{}\cPqb{}\PW{} channel due to lower efficiency for correctly identifying \cPqb{} jets using AK8 reconstructed jets with the BEST algorithm.

\begin{figure*}
\begin{center}
\includegraphics[width=0.45\textwidth]{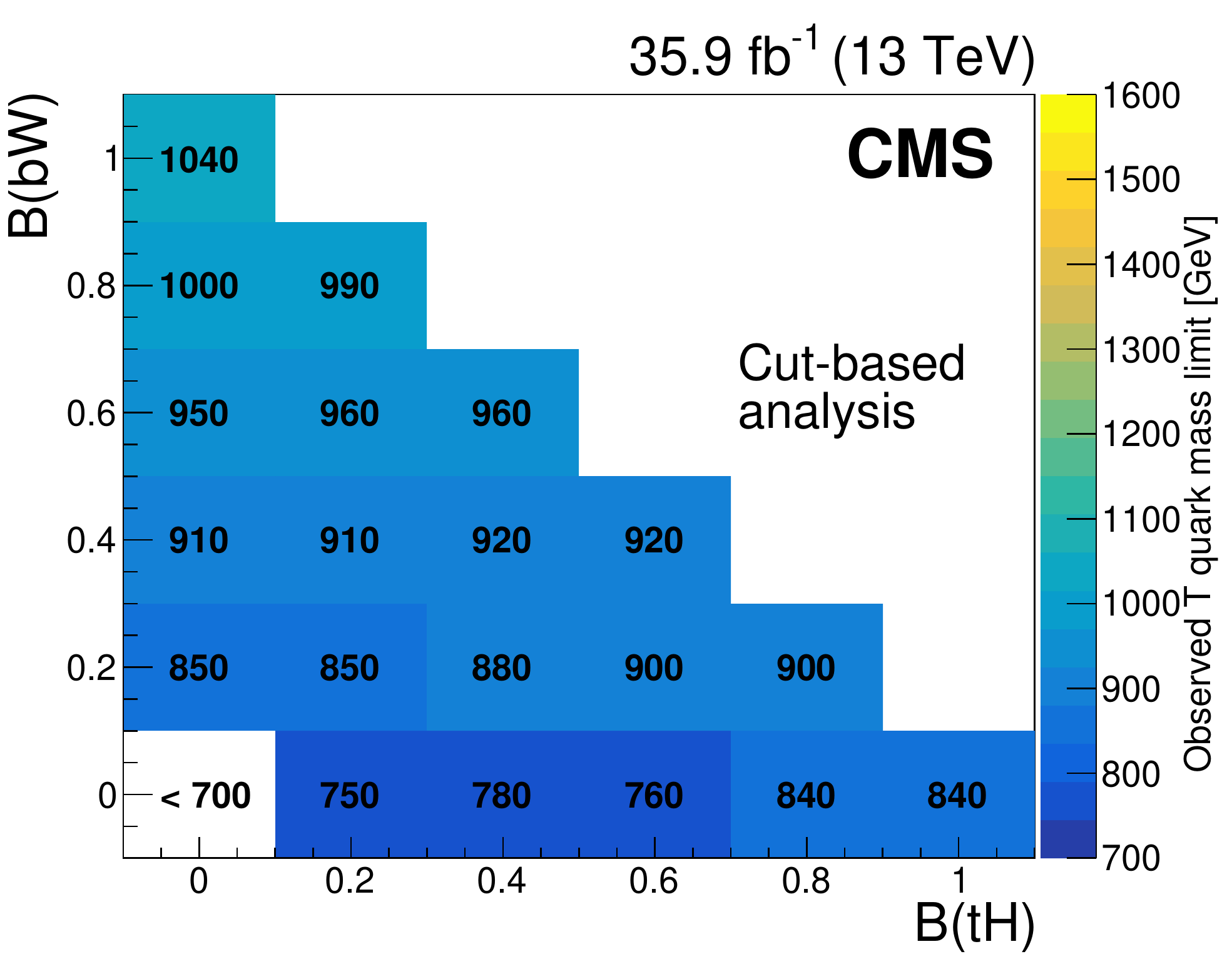}
\includegraphics[width=0.45\textwidth]{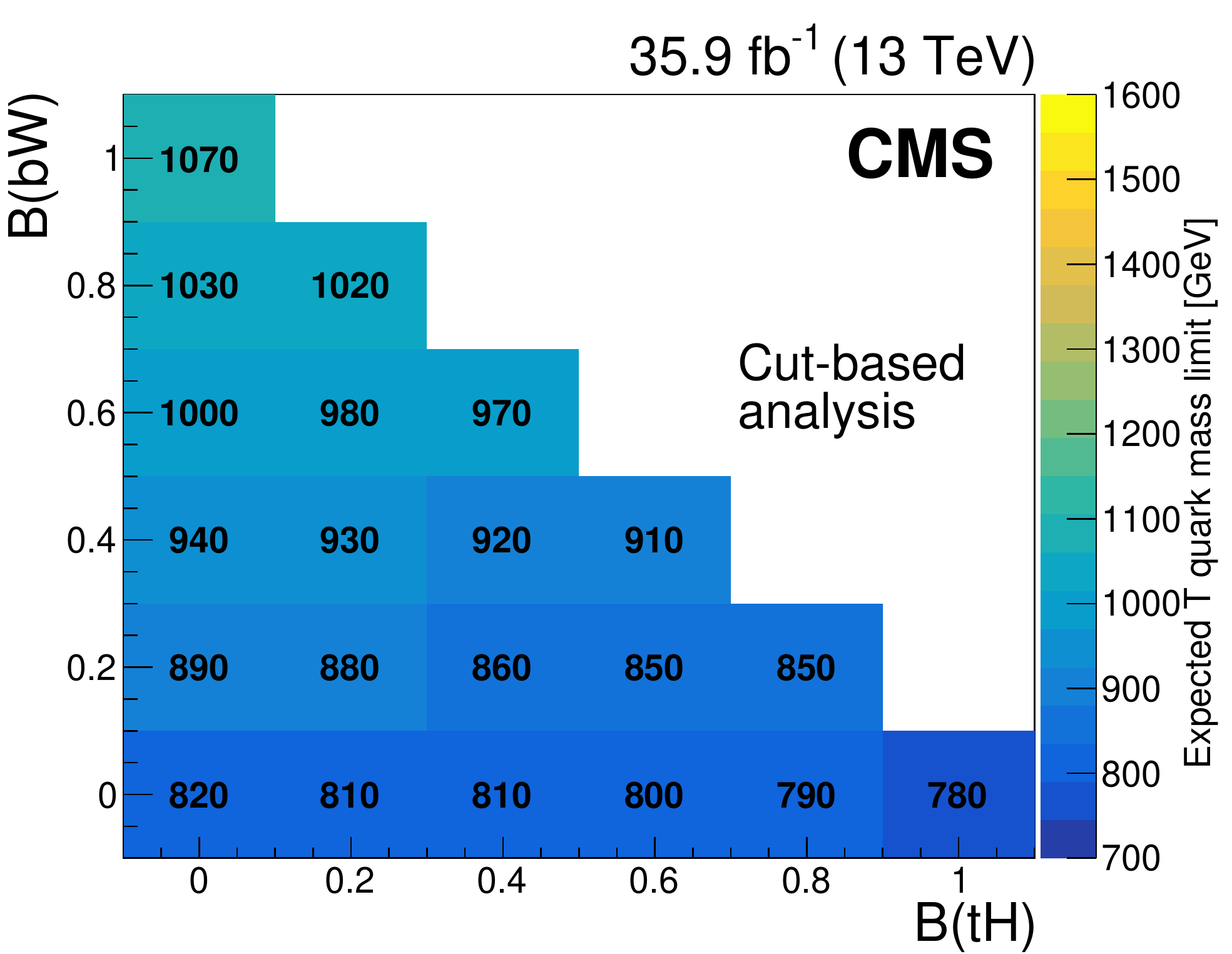}
\includegraphics[width=0.45\textwidth]{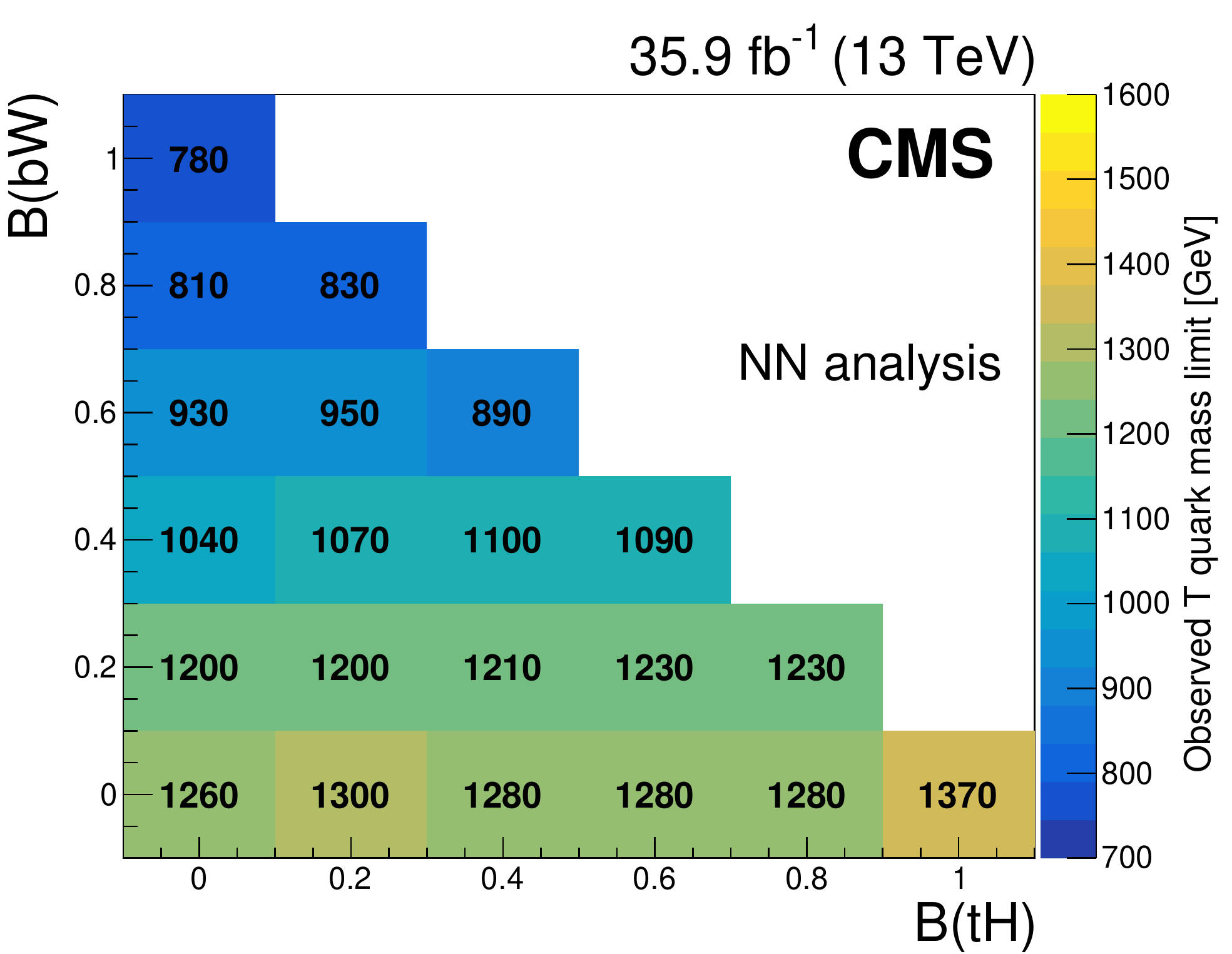}
\includegraphics[width=0.45\textwidth]{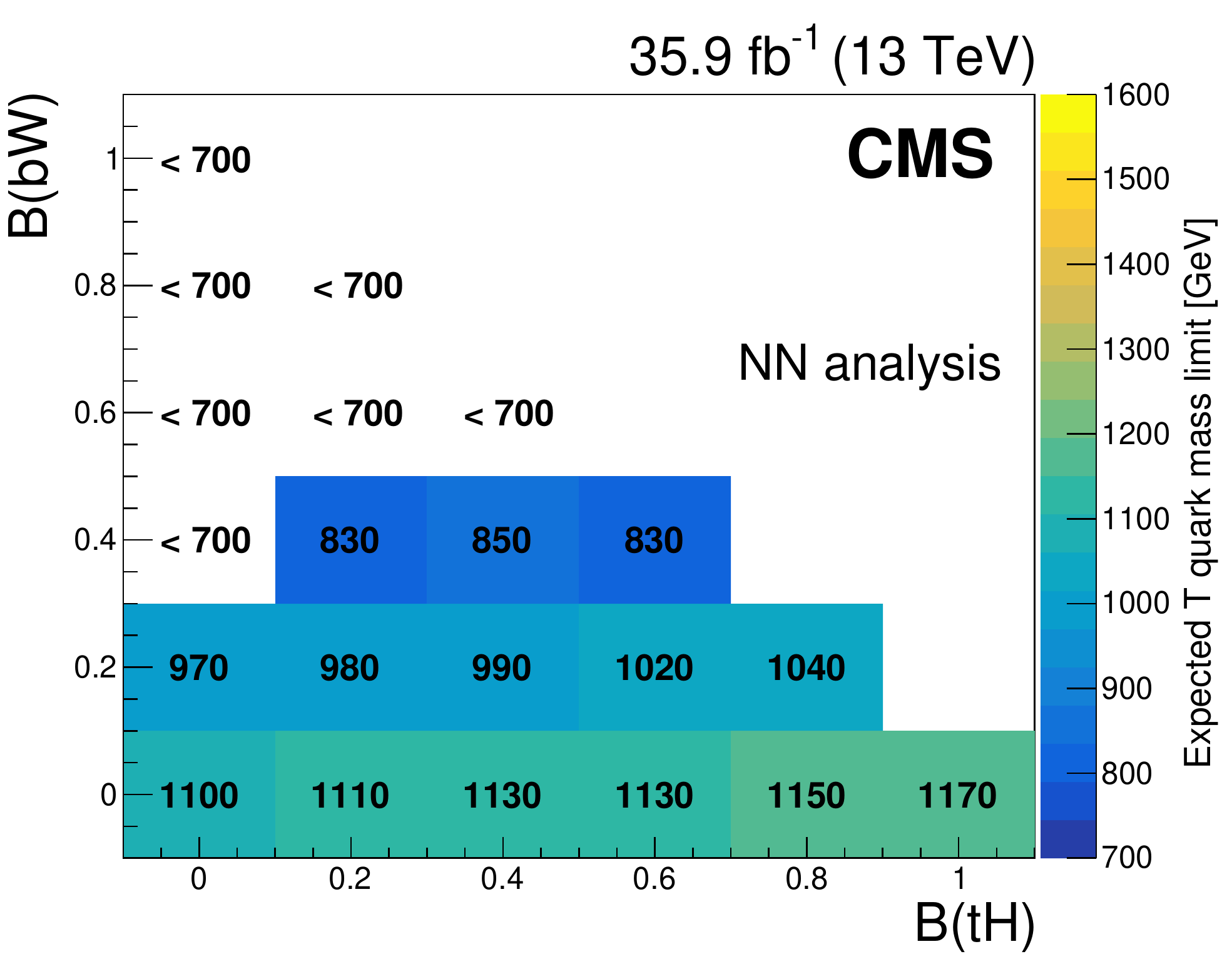}
\caption{Observed (left) and expected (right) mass exclusion limits at $95\%{}$ confidence level for each combination of \PQT quark branching fractions, in the cut-based analysis (upper) and NN analysis (lower).}
\label{fig:triangle}
\end{center}
\end{figure*}

\begin{figure*}
\begin{center}
\includegraphics[width=0.45\textwidth]{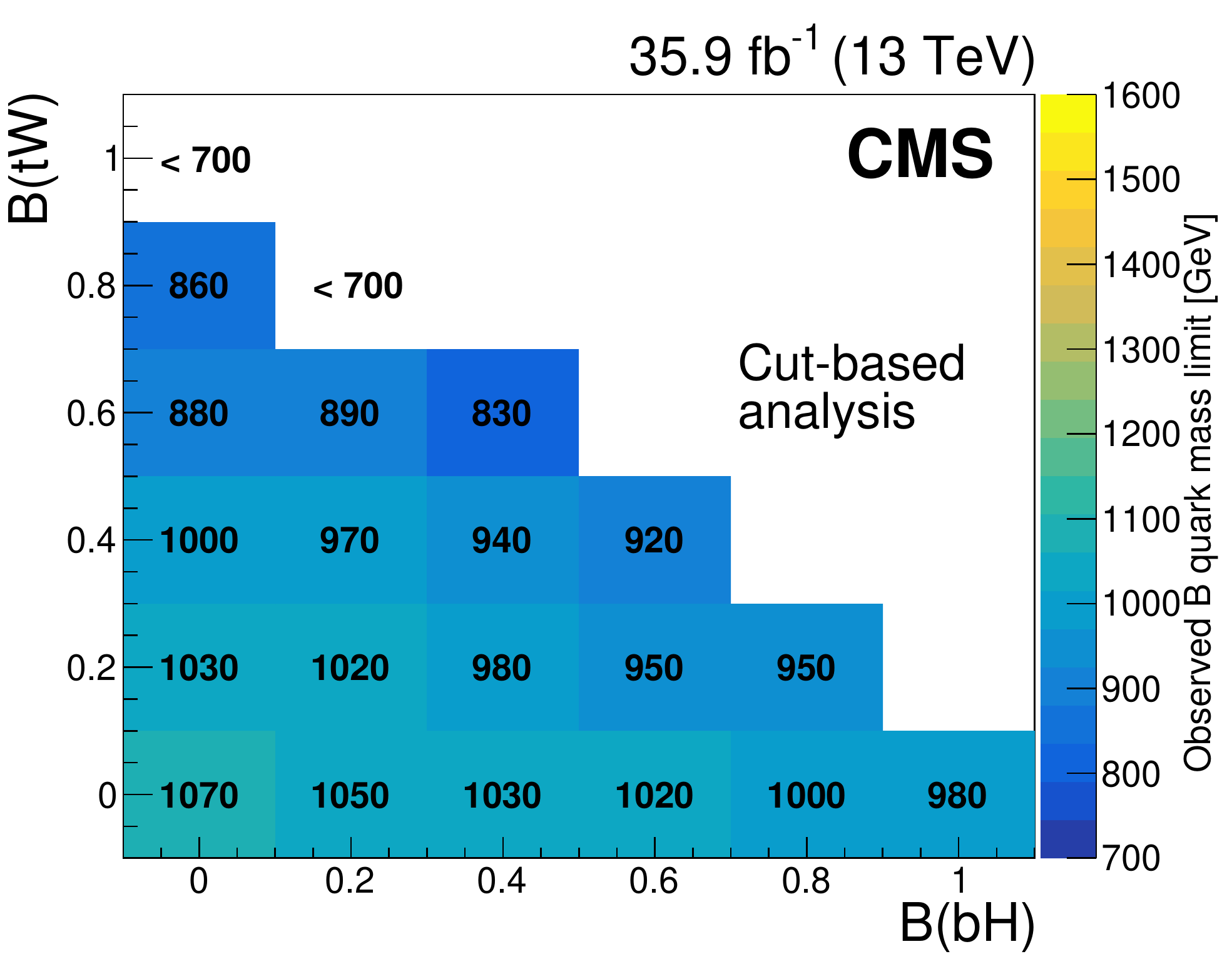}
\includegraphics[width=0.45\textwidth]{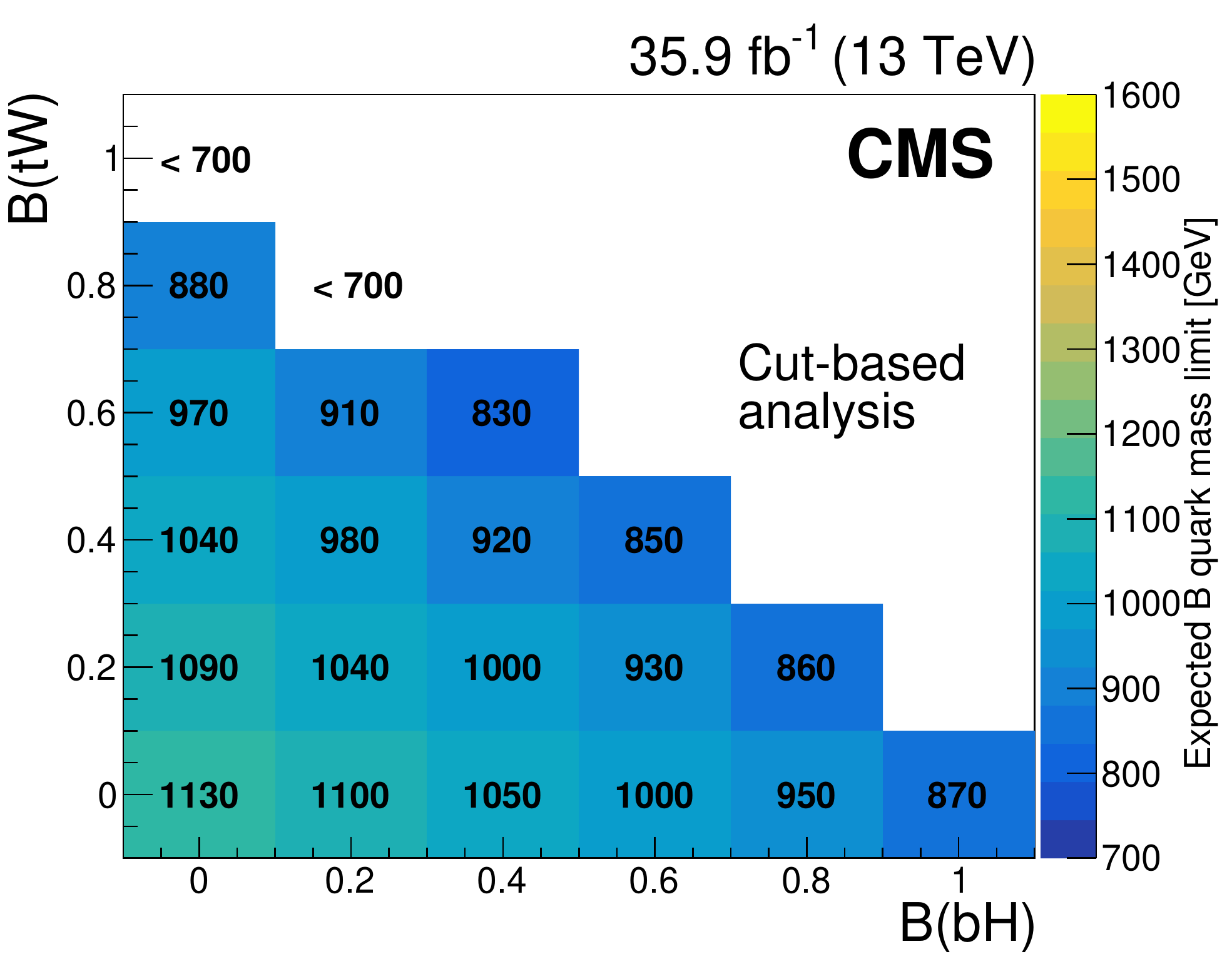}
\includegraphics[width=0.45\textwidth]{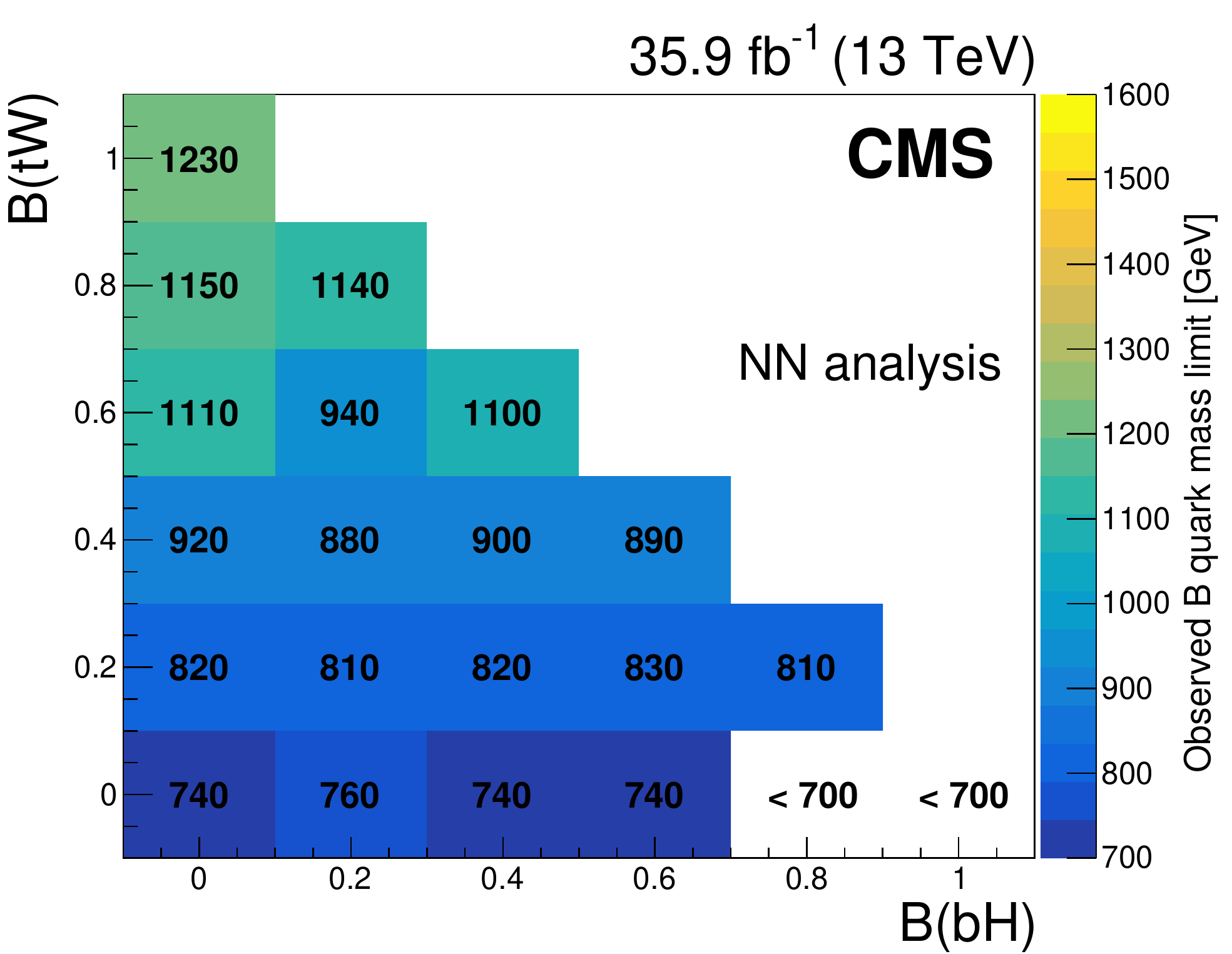}
\includegraphics[width=0.45\textwidth]{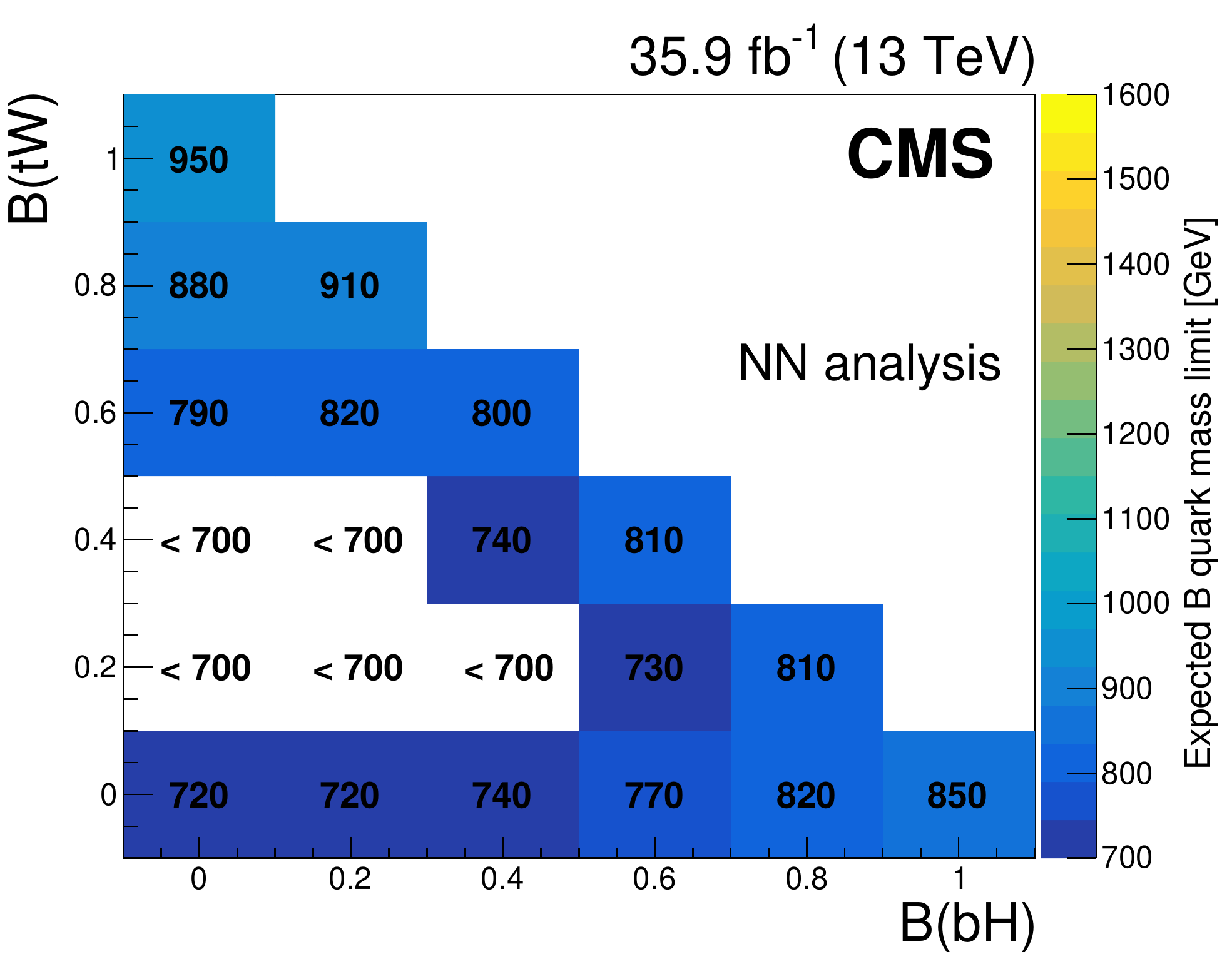}
\caption{Observed (left) and expected (right) mass exclusion limits at $95\%{}$ confidence level for each combination of \PQB quark branching fractions, in the cut-based analysis (upper) and NN analysis (lower).}
\label{fig:triangleBB}
\end{center}
\end{figure*}

A scan over all branching fractions considered is performed in increments of 0.2, with the results translated to limits on the VLQ mass.  Figure \ref{fig:triangle} shows the results for the \PQT quark graphically, with the values tabulated in Table \ref{tab:exclusion}. Figure \ref{fig:triangleBB} and Table \ref{tab:exclusionBB} show the corresponding results for the \PQB quark.

We exclude vector-like \PQT quark masses ranging from 740\GeV, up to 1370\GeV for the \cPqt{}\PH{} decay mode in the NN analysis.  The cut-based analysis provides additional sensitivity to the \cPqb{}\PW{} decay mode, with a \PQT quark mass exclusion of 1040\GeV for \PQT decays solely to \cPqb{}\PW{}.  These results complement the existing results from other decay channels, and in the hadronic channel extends the excluded \PQT quark mass from 705\GeV obtained in the previous 8\TeV analysis \cite{PhysRevD.93.012003} to 1040\GeV. For vector-like \PQB quarks, sensitivity is lost because of the additional \cPqb{} quarks present in the \PQB decays, for which the BEST analysis has a larger misidentification rate.  The cut-based analysis is not currently optimized for \PQB quarks, however does provide some complementary sensitivity to the \cPqb{}\PZ{} decay mode.  These analyses exclude vector-like \PQB quarks with masses up to 1230\GeV, for \PQB decays solely to \cPqt{}\PW{}.  A mass exclusion of 1070\GeV is obtained in the cut-based analysis for the \cPqb{}\PZ{} decay mode scenario.

\begin{table*}
\topcaption{Exclusion limits at $95\%{}$ confidence level presented in terms of the \PQT quark mass, for the different branching fraction scenarios considered, in each of the two analyses.}
\begin{center}
\begin{scotch}{ccc cccc}
 & & & \multicolumn{2}{c}{Observed limits [\GeVns{}]} & \multicolumn{2}{c}{Expected limits [\GeVns{}]} \\
$\mathcal{B}(\cPqt\PZ)$ & $\mathcal{B}(\cPqb\PW)$ & $\mathcal{B}(\cPqt\PH)$  & Cut-based & NN  & Cut-based & NN  \\
\hline
0.0 & 0.0 & 1.0 & 840 &  1370 & 780 &  1170 \\
0.0 & 0.2 & 0.8 & 900 & 1230 & 850 & 1040 \\
0.0 & 0.4 & 0.6 & 920 & 1090 & 910 &  830 \\
0.0 & 0.6 & 0.4 & 960 &  890 & 970 &   $<$700 \\
0.0 & 0.8 & 0.2 & 990 &  830 & 1020 &  $<$700 \\
0.0 & 1.0 & 0.0 & 1040 & 780 & 1070 &  $<$700 \\
0.2 & 0.0 & 0.8 & 840 &  1280 & 790 &  1150 \\
0.2 & 0.2 & 0.6 & 900 & 1230 & 850 & 1020 \\
0.2 & 0.4 & 0.4 & 920 & 1090 & 920 &  850 \\
0.2 & 0.6 & 0.2 & 960 &  950 & 980 &   $<$700 \\
0.2 & 0.8 & 0.0 & 1000 &  810 & 1030 &  $<$700 \\
0.4 & 0.0 & 0.6 & 760 &  1280 & 800 &  1130 \\
0.4 & 0.2 & 0.4 & 880 &  1210 & 860 &  990 \\
0.4 & 0.4 & 0.2 & 910 & 1070 & 930 &  830 \\
0.4 & 0.6 & 0.0 & 950 &  930 & 1000 &   $<$700 \\
0.6 & 0.0 & 0.4 & 780 &  1280  & 810 & 1130 \\
0.6 & 0.2 & 0.2 & 850 &  1210 & 880 &  980 \\
0.6 & 0.4 & 0.0 & 910 & 1040  & 940 &   $<$700 \\
0.8 & 0.0 & 0.2 & 750 &  1300 & 810 &  1110 \\
0.8 & 0.2 & 0.0 & 850 &  1210 & 890  &  970 \\
1.0 & 0.0 & 0.0 & $<$700 &  1260 & 920 &  1100 \\

\end{scotch}
\label{tab:exclusion}
\end{center}
\end{table*}

\begin{table*}
\topcaption{Exclusion limits at $95\%{}$ confidence level presented in terms of the \PQB quark mass, for the different branching fraction scenarios considered, in each of the two analyses.}
\begin{center}
\begin{scotch}{ccc cccc}
 & & & \multicolumn{2}{c}{Observed limits [\GeVns{}]} & \multicolumn{2}{c}{Expected limits [\GeVns{}]} \\
$\mathcal{B}(\cPqb\PZ)$ & $\mathcal{B}(\cPqt\PW)$ & $\mathcal{B}(\cPqb\PH)$  & Cut-based & NN  & Cut-based & NN  \\
\hline
0.0 & 0.0 & 1.0 & 980 &  $<$700 & 870 &  850 \\
0.0 & 0.2 & 0.8 & 950 &  810 & 860 &  810 \\
0.0 & 0.4 & 0.6 & 920 &  890 & 850 & 810 \\
0.0 & 0.6 & 0.4 & 830 & 1100 & 830 & 800 \\
0.0 & 0.8 & 0.2 & $<$700 & 1140 & $<$700 & 910 \\
0.0 & 1.0 & 0.0 & $<$700 & 1230 & $<$700 & 950 \\
0.2 & 0.0 & 0.8 & 1000 &   $<$700 & 950 & 820 \\
0.2 & 0.2 & 0.6 & 950 &  830 & 930 & 730 \\
0.2 & 0.4 & 0.4 & 940 &  900 & 920 &  740 \\
0.2 & 0.6 & 0.2 & 890 &  940 & 910 &  820 \\
0.2 & 0.8 & 0.0 & 860 & 1150 & 880 &  880 \\
0.4 & 0.0 & 0.6 & 1020 & 740 & 1000 & 770 \\
0.4 & 0.2 & 0.4 & 980 &  820 & 1000 &  $<$700 \\
0.4 & 0.4 & 0.2 & 970 &  880 & 980 &  $<$700 \\
0.4 & 0.6 & 0.0 & 880 & 1110 & 970 & 790 \\
0.6 & 0.0 & 0.4 & 1030 &  740  & 1050 & 740 \\
0.6 & 0.2 & 0.2 & 1020 & 810 & 1040 &  $<$700\\
0.6 & 0.4 & 0.0 & 1000 & 920 & 1040 &  $<$700 \\
0.8 & 0.0 & 0.2 & 1050 & 760 & 1100 & 720 \\
0.8 & 0.2 & 0.0 & 1030 & 820 & 1090 &  $<$700\\
1.0 & 0.0 & 0.0 & 1070 & 740 & 1130 & 720 \\

\end{scotch}
\label{tab:exclusionBB}
\end{center}
\end{table*}

\section{Summary}
\label{sec:summary}

Two independent searches for vector-like \PQT and \PQB quarks using the fully hadronic final states have been presented.  Both searches use data collected by the CMS experiment in 2016 at a center-of-mass energy of 13\TeV, corresponding to an integrated luminosity of 35.9\fbinv.  A cut-based analysis, using jet substructure observables to identify hadronic decays of boosted \PW{} bosons, targets the \cPqb{}\PW{} decay mode of the \PQT quark, and improves sensitivity relative to results of such searches conducted previously.  The analysis uses a quantum chromodynamics multijet background estimation method based on shape and rate extrapolations from various control regions to the signal region. Improvements in \PW{} tagging techniques, as well as the addition of signal regions requiring just a single \PW{}-tagged jet, enhance the performance of this analysis relative to previous searches based on different strategies.  This search extends the \PQT quark mass exclusion to 1040\GeV, relative to the previous exclusion of 705\GeV obtained by a similar analysis targeting the \cPqb{}\PW{} decay mode using data collected at 8\TeV~\cite{PhysRevD.93.012003}.

A new strategy is presented and compared with the traditional cut-based approach. The neural network analysis uses a multiclassification technique, the boosted event shape tagger algorithm, to identify jets originating from heavy objects such as \cPqt{} or \cPqb{} quarks, and \PW{}, \PZ{}, or \PH{}.  This allows the analysis to be sensitive to all decay modes of the \PQT and \PQB quarks.  Using classification fractions, the dominant multijet background is estimated using data.  The neural network analysis provides sensitivity for the \cPqt{}\PH{} and \cPqt{}\PZ{} decay modes competitive with that obtained by other searches utilizing lepton+jets or multilepton topologies.

For each analysis, results are presented in terms of cross section limits for the pair production of \PQT and \PQB quarks, along with exclusion limits in terms of the \PQT and \PQB quark masses, for the different combinations of branching fractions considered. The mass exclusion limits at $95\%{}$ confidence level for the neural network analysis range from 740 to 1370\GeV, providing comparable sensitivity to the CMS searches utilizing leptons, which exclude vector-like quark masses in the range 910--1300\GeV~\cite{Sirunyan:2018omb}.   These results represent the most stringent limits on pair produced vector-like quarks in the fully hadronic channel to date.

\begin{acknowledgments}

We congratulate our colleagues in the CERN accelerator departments for the excellent performance of the LHC and thank the technical and administrative staffs at CERN and at other CMS institutes for their contributions to the success of the CMS effort. In addition, we gratefully acknowledge the computing centers and personnel of the Worldwide LHC Computing Grid for delivering so effectively the computing infrastructure essential to our analyses. Finally, we acknowledge the enduring support for the construction and operation of the LHC and the CMS detector provided by the following funding agencies: BMBWF and FWF (Austria); FNRS and FWO (Belgium); CNPq, CAPES, FAPERJ, FAPERGS, and FAPESP (Brazil); MES (Bulgaria); CERN; CAS, MoST, and NSFC (China); COLCIENCIAS (Colombia); MSES and CSF (Croatia); RPF (Cyprus); SENESCYT (Ecuador); MoER, ERC IUT, PUT and ERDF (Estonia); Academy of Finland, MEC, and HIP (Finland); CEA and CNRS/IN2P3 (France); BMBF, DFG, and HGF (Germany); GSRT (Greece); NKFIA (Hungary); DAE and DST (India); IPM (Iran); SFI (Ireland); INFN (Italy); MSIP and NRF (Republic of Korea); MES (Latvia); LAS (Lithuania); MOE and UM (Malaysia); BUAP, CINVESTAV, CONACYT, LNS, SEP, and UASLP-FAI (Mexico); MOS (Montenegro); MBIE (New Zealand); PAEC (Pakistan); MSHE and NSC (Poland); FCT (Portugal); JINR (Dubna); MON, RosAtom, RAS, RFBR, and NRC KI (Russia); MESTD (Serbia); SEIDI, CPAN, PCTI, and FEDER (Spain); MOSTR (Sri Lanka); Swiss Funding Agencies (Switzerland); MST (Taipei); ThEPCenter, IPST, STAR, and NSTDA (Thailand); TUBITAK and TAEK (Turkey); NASU and SFFR (Ukraine); STFC (United Kingdom); DOE and NSF (USA).

\hyphenation{Rachada-pisek} Individuals have received support from the Marie-Curie program and the European Research Council and Horizon 2020 Grant, contract Nos.\ 675440 and 765710 (European Union); the Leventis Foundation; the A.P.\ Sloan Foundation; the Alexander von Humboldt Foundation; the Belgian Federal Science Policy Office; the Fonds pour la Formation \`a la Recherche dans l'Industrie et dans l'Agriculture (FRIA-Belgium); the Agentschap voor Innovatie door Wetenschap en Technologie (IWT-Belgium); the F.R.S.-FNRS and FWO (Belgium) under the ``Excellence of Science -- EOS" -- be.h project n.\ 30820817; the Beijing Municipal Science \& Technology Commission, No. Z181100004218003; the Ministry of Education, Youth and Sports (MEYS) of the Czech Republic; the Lend\"ulet (``Momentum") Programme and the J\'anos Bolyai Research Scholarship of the Hungarian Academy of Sciences, the New National Excellence Program \'UNKP, the NKFIA research grants 123842, 123959, 124845, 124850, 125105, 128713, 128786, and 129058 (Hungary); the Council of Science and Industrial Research, India; the HOMING PLUS program of the Foundation for Polish Science, cofinanced from European Union, Regional Development Fund, the Mobility Plus program of the Ministry of Science and Higher Education, the National Science Center (Poland), contracts Harmonia 2014/14/M/ST2/00428, Opus 2014/13/B/ST2/02543, 2014/15/B/ST2/03998, and 2015/19/B/ST2/02861, Sonata-bis 2012/07/E/ST2/01406; the National Priorities Research Program by Qatar National Research Fund; the Programa Estatal de Fomento de la Investigaci{\'o}n Cient{\'i}fica y T{\'e}cnica de Excelencia Mar\'{\i}a de Maeztu, grant MDM-2015-0509 and the Programa Severo Ochoa del Principado de Asturias; the Thalis and Aristeia programs cofinanced by EU-ESF and the Greek NSRF; the Rachadapisek Sompot Fund for Postdoctoral Fellowship, Chulalongkorn University and the Chulalongkorn Academic into Its 2nd Century Project Advancement Project (Thailand); the Welch Foundation, contract C-1845; and the Weston Havens Foundation (USA).

\end{acknowledgments}
\bibliography{auto_generated}

\cleardoublepage \appendix\section{The CMS Collaboration \label{app:collab}}\begin{sloppypar}\hyphenpenalty=5000\widowpenalty=500\clubpenalty=5000\vskip\cmsinstskip
\textbf{Yerevan Physics Institute, Yerevan, Armenia}\\*[0pt]
A.M.~Sirunyan$^{\textrm{\dag}}$, A.~Tumasyan
\vskip\cmsinstskip
\textbf{Institut für Hochenergiephysik, Wien, Austria}\\*[0pt]
W.~Adam, F.~Ambrogi, T.~Bergauer, J.~Brandstetter, M.~Dragicevic, J.~Erö, A.~Escalante~Del~Valle, M.~Flechl, R.~Frühwirth\cmsAuthorMark{1}, M.~Jeitler\cmsAuthorMark{1}, N.~Krammer, I.~Krätschmer, D.~Liko, T.~Madlener, I.~Mikulec, N.~Rad, J.~Schieck\cmsAuthorMark{1}, R.~Schöfbeck, M.~Spanring, D.~Spitzbart, W.~Waltenberger, C.-E.~Wulz\cmsAuthorMark{1}, M.~Zarucki
% ...
\vskip\cmsinstskip
\dag: Deceased\\
1:  Also at Vienna University of Technology, Vienna, Austria\\
% ...
\end{sloppypar}
\end{document}